\documentclass[twocolumn,aps,pra,showpacs]{revtex4}

\usepackage{graphicx}
\usepackage{amssymb}

\newcommand{\bfr}{{\bf r}}

\newcommand{\bfk}{{\bf k}}
\newcommand{\bfK}{{\bf K}}
\newcommand{\bfq}{{\bf q}}
\newcommand{\bfQ}{{\bf Q}}

\newcommand{\bfz}{{\bf 0}}

\newcommand{\hrho}{\hat{\rho}}

\newcommand{\hc}{\hat{c}}
\newcommand{\hcd}{\hat{c}^\dag}
\newcommand{\ddt}{\frac{d}{dt}}

\newcommand{\la}{\langle}
\newcommand{\ra}{\rangle}

\newcommand{\beq}{\begin{equation}}
\newcommand{\eeq}{\end{equation}}
\newcommand{\beqa}{\begin{eqnarray}}
\newcommand{\eeqa}{\end{eqnarray}}
\newcommand{\ba}{\begin{array}}
\newcommand{\ea}{\end{array}}
\renewcommand{\o}[1]{^{(#1)}}

\newcommand{\nn}{\nonumber\\}
\newcommand{\str}{^\ast}

\newcommand{\Oct}{\frac{\Omega_c}{2}}

\newcommand{\Odt}{\frac{\Omega_d}{2}}

\newcommand{\emkrj}{e^{-i\bfk\cdot\hat\bfr_j}}

\newcommand{\eKrj}{e^{i\bfK\cdot\hat\bfr_j}}
\newcommand{\emKrj}{e^{-i\bfK\cdot\hat\bfr_j}}

\newcommand{\sumk}{\sum_{\bfk}}

\newcommand{\sumj}{\sum_{j=1}^N}

\newcommand{\prodj}{\prod_{j=1}^N}

\newcommand{\sumq}{\sum_\bfq}

\newcommand{\hN}{\hat{N}}
\newcommand{\hS}{\hat{S}}

\newcommand{\hA}{\hat{A}}

\newcommand{\ha}{\hat{a}}
\newcommand{\had}{\hat{a}^\dag}

\newcommand{\hU}{\hat{U}}
\newcommand{\hUd}{\hat{U}^\dag}
\newcommand{\hV}{\hat{V}}
\newcommand{\hH}{\hat{H}}

\renewcommand{\ss}{\scriptstyle}
\newcommand{\nl}{\nonumber\\}
\newcommand{\hvr}{\hat{\varrho}}

\begin{document}


\title{Ultra-bright omni-directional collective emission of correlated photon pairs from atomic vapors}


\author{Y. P.  Huang}
\author{M. G. Moore}
\affiliation{Department of Physics \& Astronomy, Michigan State
University, East Lansing, MI  48824}


\date{\today}

\begin{abstract}
Spontaneous four-wave mixing can generate highly correlated photon
pairs from atomic vapors. We show that  multi-photon pumping of dipole-forbidden transitions
in a recoil-free geometry can result in ultra-bright pair-emission in the full $4\pi$ solid angle,
while strongly suppresses background Rayleigh scattering and associated atomic heating,
Such a system can produce photon pairs at rates of $\sim 10^{12}$ per second, given only
moderate optical depths of $10\sim100$, or alternatively, the system can generate paired photons with sub-natural bandwidths at lower production rates. We derive a rate-equation based theory of the collective atomic population and coherence
dynamics, and present numerical simulations for a toy model, as well as realistic model systems based on $^{133}$Cs and $^{171}$Yb level structures.
Lastly, we demonstrate that dark-state adiabatic following (EIT) and/or timescale hierarchy protects the paired photons from reabsorption as they propagate through an optically thick sample.
\end{abstract}
\pacs{42.65.Lm, 42.50.Ar, 42.50.Dv} 


\maketitle

\section{Introduction}

The study of correlated/entangled photon pairs has long been a
central topic in the field of quantum optics \cite{BouEkeZei00}. The
importance of paired photons is two-fold: they i) provide powerful
tools to test the peculiar aspects of quantum mechanics, such as
violations of local-realism \cite{ScuZub97,GroPatKal07,GreManEss02};
and  ii) they hold promises for advancements in quantum measurement,
communication, and information processing
\cite{BouPanMat97,MigDatSer98,
Luk03,FleImaMar05,SheKraOls06,VinMarPoo08,Kim08}. Over the past few
decades, spontaneous parametric down-conversion (SPDC) in nonlinear
crystals has been the standard source of photon pairs
\cite{HarOshBye67,KonEllFra05}. More recently, an alternative class
of biphoton sources has emerged, based on optical four-wave mixing
(FWM) in atomic vapors \cite{ZibLukScu99,VanEisAnd03,KuzBowBoo03,
BalBraKol05,KolDuBel06,ThoSimLoh06,ChaMatJen06,DuWenRu07,DuKolBel08}.
These approaches rely on collective effects \cite{Dic54,ParBarKum08} to greatly
increase the probability of correlated emission events. Compared to
SPDC, photon pairs generated via FWM in general have a much narrower
bandwidth, significantly greater temporal and spatial coherence, and
much higher conversion efficiencies. They are thus particularly
suitable for hybrid quantum communications and computations
employing atoms and photons \cite{DuaLukCir01,Kim08}, and for
high-precision quantum measurements and imaging \cite{MigDatSer98,
VinMarPoo08}.

At present, FWM photon pair sources can be categorized into three
types by level configuration. The first type, built on atomic
two-level systems, is a connected double-Rayleigh emission process
\cite{AspRogRey80,GraRigAso86}. Due to strong background Rayleigh
scattering, however, the resulting pair correlation is very weak,
without satisfying the necessary Cauchy criteria for biphoton
correlation \cite{KolDuBel06,DuWenRu07}. A second type is configured
on two-photon cascade emission in a four-level system
\cite{ChaMatJen06, ScuRay07}. While high-fidelity photon pairs are
generated, due to the unequal wavelengthes of two cascade photons,
the phase-matching condition for collective emission can only be
satisfied if the first photon is emitted by chance into a specific
small solid-angle, thus unpaired emission dominates the overall
radiation, resulting in a relatively low conversion efficiency.  The
third type employs Raman FWM (hereafter referred to as ``RFWM'') in
multilevel systems, configured on double-$\Lambda$
\cite{ZibLukScu99,BalBraKol05,KolDuBel06,DuWenRu07,DuKolBel08} or
``X'' \cite{ThoSimLoh06} level diagrams. The major challenge in
these schemes is to suppress background Rayleigh scattering, which
tends to rapidly overwhelm paired emission. Three approaches have
been proposed to for this suppression, including i) using frequency
selectors to filter out Rayleigh photons \cite{BalBraKol05}; ii)
collecting pairs along emission directions where the dipole pattern
leads to zero Rayleigh emission \cite{KolDuBel06}; and iii) using a
single-mode optical cavity to suppress Rayleigh transitions
\cite{ThoSimLoh06}. While yielding up to $10^5$ pairs per second,
all of these setups are unidirectional, where photon pairs are
produced only along certain directions. This restricts the
obtainable beam brightness of the photon pairs, since in each
momentum mode, the time separation between pairs must be
large relative to the correlation time. Lastly, in
aforementioned FWM schemes where unpaired emissions dominate, atomic
samples are rapidly thermalized due to random atomic recoils,
limiting applications of these schemes to `hot' vapors only.

Background Rayleigh scattering occurs when there is a spontaneous
one-photon channel by which atoms can return to the initial internal hyperfine level
without completing the desired biphoton emission cycle. We propose
to eliminate this in RFWM by replacing the single pump
laser with a multi-photon pump process that drives a one-photon
dipole-forbidden transition. The second part of our proposal is to
use `recoil-free' pumping, meaning that the k-vectors of all of the
driving fields sum to zero. Phase matching, which is enhanced by
collective effects, will then be satisfied whenever the two paired
photons have equal and opposite momenta, allowing paired emission
into the full $4\pi$ solid angle. As a result, the biphoton emission
rate will be enhanced by up to four orders of magnitude over
unidirectional approaches. Furthermore, since Rayleigh scattering
has been eliminated, atomic thermalization will be strongly
suppressed. Thus this scheme can also be applied in ultracold vapors
including Bose-Einstein condensates (BECs), for which it might also
be viewed as a novel in-situ non-demolition imaging technique.

A brief organization of this paper is as follows. In section
\ref{Sec.BM}, we study a `butterfly' biphoton protocol using a
simplified model, which is a viable simplification of realistic
configurations. In section \ref{Sec.RM}, we give two realistic
implementations, one employing the $399$nm-line transition in
Ytterbium (section \ref{Sec.RM.SA}), and the other using the $852$nm
D2-line transition in Cesium (section \ref{Sec.RM.CA}).  In section \ref{Sec.RA}, we then
address the critical issue of how to avoid reabsorption as the
 photon pairs propagate in an optically thick sample. This is
followed by a discussion and conclusions in section \ref{Sec.Conc}.

\section{Basic Model}
\label{Sec.BM}

In this section, we will use a simplified `toy model' to show the
important physics of the butterfly scheme for ultra-bright
photon-pairs. We present the schematic model in section
\ref{Sec.BM.TLS}, and employ a set of rate equations to solve for
atomic dynamics in section \ref{Sec.BM.PD}. Then, we examine the
time and polarization correlation of generated photon pairs in
\ref{Sec.BM.PPC} and \ref{Sec.BM.PE}. In section \ref{Sec.BM.CD}, we
estimate the threshold temperature of the present biphoton source.

\subsection{Toy Level Scheme}
\label{Sec.BM.TLS}

\begin{figure}[htp]
\includegraphics[width=8.0cm]{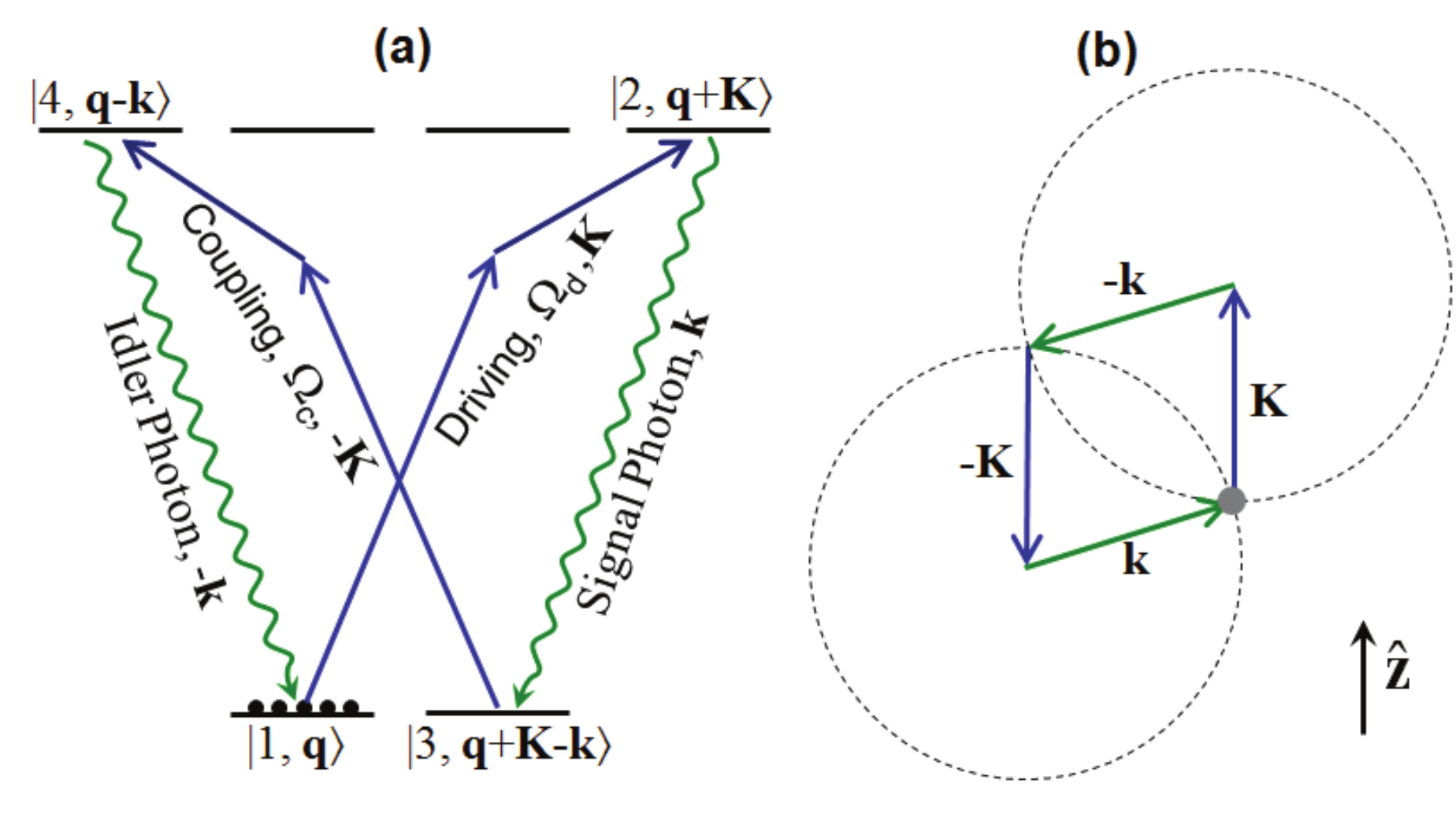}
 \caption{(Color online) A schematic model of the butterfly scheme.
 Figure (a) draws the simplified level diagram, employing multi-photon
 driving and coupling pumps.  The notation $|j,\bfk\rangle$
 indicates a single atom state in internal level $|j\rangle$ and with momentum $\hbar
 \bfk$. The momentum $\hbar\bfK$ is the net momentum of the multiphoton transition from levels $|1\ra$ to $|2\ra$.
 Figure (b) shows how a phase-matched
  diamond-like cycle in atom-recoil momentum space exists
  for any signal photon emission direction.
 \label{fig1}}
\end{figure}

A schematic level diagram of the butterfly scheme is shown in Fig.
\ref{fig1} (a). While greatly simplified with respect to a realistic
level-scheme, this model will serve to illustrate the important
dynamical effects. The physical mechanism is best illustrated in a quantum trajectory picture, as follows. A sample of atoms
initially pumped into the $|1\rangle$ state, is first weakly coupled to the excited $|2\rangle$ level via a multi-photon pump process. This imparts a net recoil momentum of $\hbar{\bf K}$, so that for an initial momentum
$\hbar\bfq$, an atom excited to $|2\rangle$ has a momentum of
$\hbar (\bfq+{\bf K})$. This excited atom will then spontaneously
decay to $|3\rangle$, emitting a `signal' photon with a random
momentum $\hbar \bfk$, shifting the atom's momentum to $\hbar
(\bfq+{\bf K}-\bfk)$. Decay from $|2\ra$ back to state $|1\rangle$, which would
generate background Rayleigh scattering, is forbidden by dipole
selection-rules. The atom in state $|3\rangle$ is then rapidly
repumped to $|4\rangle$ by a strong multi-photon coupling process.
The coupling fields are arranged to yield a net momentum of $-\hbar
{\bf K}$, leading to a momentum of $\hbar (\bfq-\bfk)$ for the
atom. From $|4\rangle$, the atom decays back to the $|1\rangle$
state, emitting an `idler' photon. Collective effects will strongly
enhance the scattering probability if the atom can be returned to
its initial momentum state $\hbar\bfq$, which will result in the
idler photon being emitted with momentum $-\hbar\bfk$.

The collective enhancement mechanism can be understood by noting
first that the emission of the signal photon with momentum
$\hbar\bfk$ imprints `which atom' information onto the atomic
ensemble via atomic recoil, provided of course that the
single-photon recoil momentum is larger than the momentum coherence
length of the sample (which for a thermal gas of free particles is
the inverse sample length). If the idler photon is then emitted with
phase-matched momentum, $\hbar\bfk_i=-\hbar\bfk$, then the atom is restored to its initial momentum state of $\hbar\bfq$, thus `erasing' the `which
atom' information, so that many-body interference enhances the emission
rate by a factor $N$. The `which atom' information will be effectively erased only when the magnitude of the
momentum difference between the initial and final states  is less than or equal to the momentum coherence length of the atom, $|\bfk+\bfk_i|\lesssim k_{coh}$, which leads to a collective emission solid-angle of $\Omega_{coh}\sim \pi (k_{coh}/k)^2$.
For a lone atom, and neglecting the dipole emission pattern, the probability of correlated emission, i.e. the probability of
$\bfk_i$ falling within $\Omega_{coh}$ of $-\bfk$, is $P_c(1)=\Omega_{coh}/4\pi$ with a non-correlated probability $P_{nc}(1)=1-\Omega_{coh}/4\pi$.
For a sample of $N$ atoms, the differential probability inside $\Omega_{coh}$ is enhanced by $N$ relative to the differential probability outside $\Omega_{coh}$, leading to a collectively enhanced correlated emission probability of $P_c(N)=N\Omega_{coh}/(N\Omega_{coh}+4\pi-\Omega_{coh})$. The probability of non-correlated emission is correspondingly reduced by unitarity to $P_{nc}(N)=(4\pi-\Omega_{coh})/(N\Omega_{coh}+4\pi-\Omega_{coh})$. Under the condition $N\Omega_{coh}/4\pi\gg 1$, the probabilities are then given approximately by $P_c(N)=1-4\pi/N\Omega_{coh}\sim 1$ and $P_{nc}(N)\approx 4\pi/N\Omega_{coh}\ll 1$. For a thermal sample of dimension $L$, we have $k_{coh}\sim 1/L$, so that $\Omega_{coh}\sim \pi/(kL)^2$. This gives a correlated emission probability of $P_c=1-1/D$, where $D=N/(2kL)^2=n\lambda^2L/16\pi^2$ is the optical depth of the sample. Other considerations aside, we clearly see that strong photon pair-correlations in $k$-space are achieved by maximizing the optical thickness of the sample.

The fact that the driving and coupling fields have zero net momenta
makes the scheme in a limited sense `recoil-free', so that phase-matched collective emission can occur
regardless of which direction the signal photon randomly `chooses'.
This is illustrated in Fig. \ref{fig1} (b), where for an arbitrarily chosen
$\bfk$, the atomic dynamics in the space of recoil momentum
undergoes a closed, diamond-like cycle. This distinguishes the
present setup from competing biphoton protocols employing
non-counter-propagating driving and coupling, where pair emissions
are restricted to be in a plane perpendicular to the nonzero net
momentum \cite{ChaMatJen06, ScuRay07,BalBraKol05,KolDuBel06}.

\subsection{Population Dynamics}
\label{Sec.BM.PD}

To study the system's dynamics, we quantize the atomic center-of-mass motion, onto the eigenmodes, $\{|\bfq\ra\}$, of a box of dimension $L$, where $L$ is the ensemble dimension. As the box traverse time, $L/v$, is long compared to the relevant dynamic timescale, $(\Gamma)^{-1}$,
we can safely impose periodic boundary conditions, so that the allowed $\bfq$ values lie on a three-dimensional cubic lattice with spacing $2\pi/L$. Eliminating the scattered light-field via the Markoff approximation, and taking into account the exchange-type symmetry respected by the Hamiltonian and the initial conditions, allows us to derive a set of rate equations for the atomic population dynamics. The advantage of rate equations, as opposed to a mean-field approach,
is that spontaneous decay is incorporated via the usual $(N+1)$ factors,  thus avoiding the need for
noise operators. For the toy model depicted in Fig. 1 (a), the rate equations are:
\begin{eqnarray}
\label{ndy}
    & & \!\!\!\!\!\!\!\!\!\! \ddt N_1 =
    \frac{i}{2}\left(\Omega_d\varrho_{21}-c.c\right)+\sum_\bfq
    \Gamma_4 \beta_{\bfq 4} N_{\bfq 4} (N_1+1), \\
    \label{ndy1a}
    & & \!\!\!\!\!\!\!\!\!\! \ddt N_2 =
    -\frac{i}{2}\left(\Omega_d\varrho_{21}-c.c\right)-\sum_\bfq
    \Gamma_2 \beta_{\bfq 2} N_2 (N_{\bfq 3}+1), \\
    & & \!\!\!\!\!\!\!\!\!\! \ddt \varrho_{21}=i \frac{\Omega_d}{2} (N_1-N_2) \nn
    & & ~~~~~~+\frac{1}{2}\varrho_{21}
      \sum_\bfq\left[\Gamma_4\beta_{\bfq 4} N_{\bfq 4}-
       \Gamma_2 \beta_{\bfq 2} (N_{\bfq 3}+1) \right], \\
    & & \!\!\!\!\!\!\!\!\!\! \ddt N_{\bfq 3} = \frac{i}{2} \left(\Omega_c\varrho_{\bfq 43}-c.c\right)+
    \Gamma_2 \beta_{\bfq 2} N_2 (N_{\bfq 3}+1), \\
    & & \!\!\!\!\!\!\!\!\!\! \ddt N_{\bfq 4} =
    -\frac{i}{2}\left(\Omega_c \varrho_{\bfq 43}-c.c\right)-
    \Gamma_4 \beta_{\bfq 4}N_{\bfq 4} (N_1+1), \nn
    & & ~~~~~~-\Gamma_4(1-\beta_{\bfq4})N_{\bfq4}\\
    \label{ndy1}
     & & \!\!\!\!\!\!\!\!\!\! \ddt \varrho_{\bfq 43}=i \frac{\Omega_c}{2} (N_{\bfq 3}-N_{\bfq 4}) +\frac{1}{2}\varrho_{\bfq 43}\times \nl
    & & ~~~~~~
      \left[\Gamma_2\beta_{\bfq 2} N_2-
       \Gamma_4 \beta_{\bfq 4}( N_1+1)-\Gamma_4(1-\beta_{\bfq4})\right],
\end{eqnarray}
a detailed derivation of these equations, as well as precise definitions of the variables, is presented in appendix \ref{Sec.APP.1}.
Oversimplifying slighty, we can think of $N_1$ and $N_2$ as the populations of state $|1\rangle$ and
$|2\rangle$, respectively, with $\varrho_{21}$ being the corresponding coherence operator.
$N_{\bfq 3}$ is the expectation number of atoms collectively excited
into state  $|3\rangle$ via emission of a signal photon with $\bfk\approx-\bfq$. Similarly, $N_{\bfq 4}$ is the
expectation number of these atoms transferred to state
$|4\rangle$ by the coupling laser, while $\varrho_{\bfq 4 3}$ is the
coherence between these two collective states. Lastly, $\Omega_d$ and
$\Omega_c$ are the effective Rabi frequencies of the driving and
coupling transitions. The spontaneous emission rates for $|2\rangle
\rightarrow |3\rangle$ and $|4\rangle\rightarrow |1\rangle$ decays, are $\Gamma_2$ and
$\Gamma_4$, respectively, and
$\beta_{\bfq 2}$ and $\beta_{\bfq 4}$ are the branching ratios for emission into the coherent-emission solid angle $\Omega_{coh}=\pi/(kL)^2$, with respect to the $\pm\bfq$ directions. For a spherical
sample of radius $L/2$, a careful calculation \cite{MooMey99} gives
\begin{equation}
\label{Eq.BM.mu}
    \beta_{\bfq
    \mu}=(1-|\hat{q}\cdot\hat{d}_\mu|^2)\frac{3}{8\pi}\left(\frac{\lambda}{L}\right)^2;\quad \mu=2,4;
\end{equation}
where $\hat{d}_2$, and $\hat{d}_4$ are the unit vectors along
the dipole moments of the $|2\rangle \rightarrow |3\rangle$ and
$|4\rangle\rightarrow |1\rangle$ transitions. Aside from dipole-emission factor, $(1-|\hat{q}\cdot\hat{d}_j|^2)$, this is in good agreement of our initial estimate $\Omega_c/4\pi\sim \lambda^2/(4\pi L)^2$.
The quantity $D_{\bfq4}=N_1\beta_{\bfq 4}$ is then the optical depth along the $\pm\bfq$ directions, with respect to the $|1\ra\leftrightarrow|4\ra$ transition.

Strictly speaking, in the toy model, atoms which spontaneously decay
from $|4\rangle$ to $|1\rangle$ by emitting rogue photons can still
participate in the next-round pair-emission cycle, as the initial
momentum of the atoms in state $|1\ra$ is irrelevant to the collectivity (in the Doppler-free regime). This recycling
process can be included in the rate-equation model by inserting a
re-feeding term $\sum_{\bfq} \Gamma_4 (1-\beta_{\bfq 4})
N_{\bfq 4}$ to equation (\ref{ndy}). In this case, the total atom
number $N_1+N_2+\sum_{\bfq} N_{\bfq3}+N_{\bfq4}$ will be conserved, while still correctly
describing the emission of rogue (uncorrelated) photons. In
realistic schemes, however, atoms can also spontaneously decay to
other ground levels which are not shown in figure \ref{fig1}, and/or
they can decay from intermediate pumping levels to $|3\rangle$. These
atoms will not be able to participate in further collective emission
cycles, unless the are somehow repumped back to $|1\rangle$.
Thus to avoid overly optimistic predictions,
we have chosen not include the recycling process in the
rate-equation model. It is noted that by completely
excluding the recycling process, the pair generation rate is
\emph{underestimated}, as a fraction of rogue-photon emitting atoms will always be recycled.
On the other hand, atoms which
decay from $|4\ra$ into levels other than $|1\rangle$ may be repeatedly re-excited to other levels by the pumping fields, and
thus emit additional rogue photons. If such rogue photons can not be
filtered out, they will contribute to the impurity of the collected
biphoton beams. For present, however, we only focus on the
short-time behavior of the system, up to a point when the atom loss
is about $10\%$. In this time interval, the present dynamical
model with rate equations (\ref{ndy})-(\ref{ndy1}) is reasonably valid.

The total emission rates for signal and idler photons, corresponding
to the (enhanced) decay rates of the $|2\rangle$ and $|4\rangle$
levels, are
\beqa
R_S&=&\Gamma_2 \sum_\bfq \beta_{\bfq 2} N_2(N_{\bfq 3}+1),\\
 R_I&=&\Gamma_4 \sum_\bfq \beta_{\bfq 4} N_{\bfq 4}(N_1+1),
 \eeqa
respectively. Assuming steady-state, clearly we must have $R_I\le
R_S$. If $R_I<R_S$, more signal photons are generated than idler
photons, so that pairing is weak. Thus, at a minimum, strong pairing
requires $R_I=R_S$. Focusing on a single $\bfq$ mode, and
assuming $\beta_{\bfq 2}=\beta_{\bfq 4}$ and $\Gamma_2=\Gamma_4$, we
find
\beq
\frac{R_I(\hat{k})}{R_S(\hat{k})}=\left(\frac{N_1+1}{N_2}\right)\left(\frac{N_{\bfq4}}{N_{\bfq 3}+1}\right).
\label{RIoverRS}
\eeq
If we assume a strong drive, $\Omega_d\gtrsim
\Gamma_2$, we have $N_2\sim N_1$, which means we must also have
$N_{\bfq 4}\sim N_{\bfq 3}\gg 1$, which in turn requires a strong
coupling field, $\Omega_c\gtrsim \Gamma_4\beta_{\bfq 4}N_1$. We find,
however, that dynamically this approach doesn't work, as it leads to
a build-up of population in $N_{\bfq 3}$ and $N_{\bfq 4}$ without
strong pairing. This leaves the case of weak driving, $\Omega_d\ll
\Gamma_2$ so that $N_2\ll N_1$. This then requires $N_{\bfq
4}=\frac{N_2(N_{\bfq 3}+1)}{N_1}$, which can be arranged by
adjusting the drive and coupler strengths and detunings, and
provided $N_{\bfq 4}\ll 1$, results in strong pairing
\cite{KolDuBel06,DuKolBel08}.

The impurity of collected biphoton beams in the present toy-model
comes from spontaneous emission of rogue idler photons into non
phase-matched angles. The total emission rate of rogue photon is
given by $R_{rogue}=\Gamma_4 \sum_\bfq N_{\bfq 4}(1-\beta_{\bfq
4})\approx \Gamma_4 \sum_\bfq N_{\bfq 4}$, where
$\beta_{\bfq 4}\ll 1$. The ratio of paired idler to rogue idler
photons is then given by $R_I/R_{rogue}=(N_1+1) \bar{\beta}$,
where $\bar{\beta}=\sum_\bfq \beta_{\bfq 4} N_{\bfq 4}/\sum_\bfq
N_{\bfq 4}$ is the mean collectivity averaging over emission
angles. For a spherical cloud of radius $R$, this is roughly
$\lambda^2/16 \pi R^2$. Defining the optical depth of a spherical
cloud as $D=N_1\bar{\beta}$, the pair to rogue ratio is simply
$R_I/R_{rogue}=D$, i.e., the optical depth. For typical samples of
$D\sim 100$, there is then about one rogue photon per 100 pairs.

In general, the emission of a rogue photon leads to heating of the sample, due to random non-zero net-recoil. In a unidirectional scheme,
the vast majority of photons are not paired, so that heating occurs at the usual single-photon decay spontaneous heating rate. Such schemes are thus only applicable to samples well above the recoil temperature. With the omnidirectional approach, the majority of spontaneous photons come in correlated pairs and thus impart no recoil kick. The heating rate is then reduced by a factor of the optical depth $D$, which should allow interesting experiments to be performed at or below the recoil temperature.
For example, if a  BEC of $N$ atoms is used as an omnidirectional biphoton source, the condensate
only depletes at a rate of $R_{rogue}$, so that, e.g. $N D/10 \approx 10 N$ photon
pairs could be generated with only $10\%$  of the condensate atoms being lost.
This means that if desired, the present butterfly scheme can be
used to directly  image condensates \emph{in situ}, in a relatively nondestructive manner. In other words, the BEC would exhibit resonance fluorescence, but  with strongly suppressed heating. The additional brightness might, e.g., yield improved atom-number estimation.

Similarly to the case of two-photon cascade emission \cite{ScuZub97},
here the bandwidth of both signal and idler photons in the strong
coupling regime ($|\Omega_c|\gg D\,\Gamma_4$)  is given by
$\Gamma_4 D/2$, where $\Gamma_4 D$ is the superradiance broadened linewidth of state $|4\ra$. This is because the intermediate
$|3\rangle$ and $|4\rangle$ levels participate the pair emission by
first forming dressed states of $|\pm \rangle=(|3\rangle\pm
|4\rangle)/\sqrt{2}$. The process of pair emission depicted in
figure \ref{fig1} is then physically equivalent to cascade emission
from $|2\rangle$ to $|1\rangle$, through either $|+\rangle$ or
$|-\rangle$ states. In further analogy to cascade emission, the frequency-sum of the signal and idler
photons has a much narrower linewidth of $|\Omega_d|^2/\Gamma_2$, as
given by the reciprocal of the pair emission time.
In the weak coupling regime,
$|\Omega_c|\ll D\,\Gamma_4$, the coupling dynamics is
overdamped, so that the $|4\rangle$ level can be adiabatically
eliminated. Then the pair emission is effectively through the
$|3\rangle$ level only, the linewidth of which is broadened from zero to
$|\Omega_c|^2/D\,\Gamma_4$ by the coupling laser. This,
together with the fact that we choose
$|\Omega_c|^2/D\,\Gamma_4\gg| \Omega_d|^2/\Gamma_2$
for a strong pairing effect, gives a bandwidth of
$\frac{|\Omega_c|^2}{D\,\Gamma_4}\ll D\,\Gamma_4$ for the emitted
photons. In the very weak coupling limit, with
$|\Omega_c|<\sqrt{D}\Gamma_4 $, this means that correlated photon pairs of sub-natural bandwidth can be
generated.

To study the performance of this system, we numerically solve
equations (\ref{ndy})-(\ref{ndy1}).  One question that must be decided is how-many and which box eigenmodes need to be included in the simulation.
Firstly, we note that in momentum space, the width of the energy shell for the signal photons is $\Delta k=\Gamma_4 D/c\sim 1\mbox{m}^{-1}$, while the size of a single eigenmode is $1/L$. Thus for normal sample sizes of $L\ll 1\mbox{m}$, only a single shell of modes of radius $2\pi/\lambda$ participates in the dynamics, where $\lambda\sim 10^{-7}\mbox{m}$ is the wavelength of the $|2\ra\to|3\ra$ transition. This leads to a mode number of $M=4\pi/\Omega_c\sim(L/\lambda)^2$, which could range anywhere from $10^4$ to $10^{10}$, depending on the sample size.
To handle the dipole pattern, we then
sort these modes $\{\bfq\}$ into $15$ groups. The $j$-th
($j=1,2,...15$) group contains those quasi-modes satisfying
$\theta_\bfq\in [\theta_j, \theta_j+\pi/30)$, where  $\theta_j=(j-1)\pi/30$, and
$\theta_\bfq\in(0,\pi/2]$ is the angle between $\bfq$ and
the $\hat{d}_2$ axis.
Modes in the $j^{th}$ group, are assigned with mean collectivity
parameters, $f_{j2}$, and $ f_{j4}$, obtained by averaging over modes
inside the group. In this approximation, all quasi modes within the
same group will yield identical dynamics, so that only
of one mode in each group needs to be included in the dynamical model.
The effects of the other modes can then be included by weighting each
representative mode by the number of modes on the interval  $[\theta_j-\pi/30, \theta_j)\sim(L/\lambda)^2\sin(j\pi/30)$.
In this way, we are able to reduce the number of coupled rate equations from $10^4$-$10^{10}$, down to fewer than 100, while still incorporating the effects of the dipole radiation pattern.

\begin{figure}[htp]
\includegraphics[width=8.9cm]{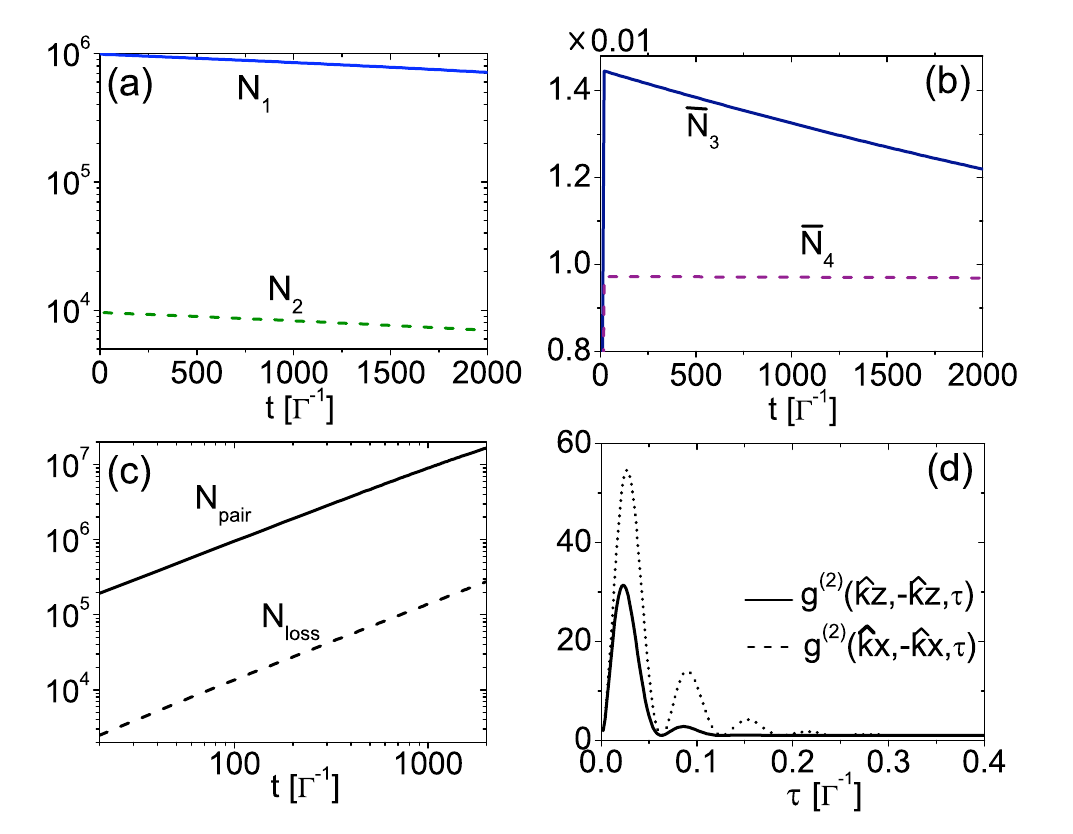}
 \caption{(Color online) Figures (a)-(b) show the evolutions of
 $N_1,N_2,\overline{N}_3, \overline{N}_{4}$, while (c) compares the number of generated
 photon pairs $N_{pair}$ and lost atoms $N_{loss}$.
 Figure (d) plots the second-order correlation function
 $g^{(2)}(\bfk,-\bfk,\tau)$ for $\bfk=k\hat{z}$ and $\bfk=k\hat{x}$,
 respectively. Parameters are given in text.
 \label{fig2}}
\end{figure}

We consider an example of $N=10^6$ atom, with a
spherically-symmetric Gaussian density distribution of radius
$L=34\lambda$, corresponding to an optical depth of $D=70$. We
consider resonant driving and coupling pumps propagating along
$\hat{z}$ and $-\hat{z}$ directions, and take
$\Gamma_2=\Gamma_4\equiv\Gamma$, $\Omega_d=0.1 \Gamma$,
and $\Omega_c=100\Gamma$. The results are shown in Fig. \ref{fig2}
(a)-(d). In figure (a), we plot the time evolution of $N_1, N_2$,
showing that at all times only a small fraction of atoms are
excited, i.e. $N_2\ll N_1$, as required. In figure (b), we plot the
mean atomic collective excitation numbers $\overline{N}_{3}$ and
$\overline{N}_{4}$, obtained by averaging over $\bfq$.  Both are
found to be of order of $0.01$, so that there is negligible overlap
between subsequent pairs in a given mode. The total
photon-pair number and lost atom number are shown in figure (c),
where they are found to increase linearly in time with fitted rates
of $8.3\times 10^3 \Gamma$ and $1.4 \times 10^2 \Gamma$.

\subsection{Photon Pair Correlation}
\label{Sec.BM.PPC}

To estimate the time correlation of the two photons, we calculate
the time-averaged second-order correlation function
$g^{(2)}(\bfk,-\bfk,\tau)$, defined as \cite{ScuZub97}
\begin{eqnarray}
\label{eq.g2}
    & & g^{(2)}(\bfk,-\bfk,\tau)=\frac{1}{T} \int_0^T dt \times \nl
       & &~~~~~~~~
     \frac{\langle \ha^\dag_{\bfk s}(t)\ha^\dag_{-\bfk i} (t+\tau)
     \ha_{-\bfk i}(t+\tau)\ha_{\bfk s}(t)\rangle}
     {\langle \ha^\dag_{\bfk s}(t)\ha_{\bfk s}(t)\rangle\langle
     \ha^\dag_{-\bfk i}(t+\tau)\ha_{-\bfk i}(t +
     \tau)\rangle},
\end{eqnarray}
where $\ha_{\bfk s}$ and $\ha_{\bfk i}$ are the annihilation
operators for signal and idler photons, which may or may not differ
in polarization for a given $\bfk$, and $T$ is the averaging window.
The correlation function is evaluated by first using adiabatic
following to write the photon operators in terms of the atomic
operators, where
\begin{eqnarray}
    \ha_{\bfk s}(t)&=& \ha_{\bfk s}(0)e^{-i\omega_\bfk t}-
    i g_\bfk
    \frac{\sin[(\omega_s-\omega_\bfk)t/2]}{(\omega_s-\omega_\bfk)t/2}
    \nonumber \\
     & & \times e^{-i(\omega_s+\omega_\bfk)t/2} \hat{\sigma}_{2,\bfk3}, \\
    \ha_{\bfk i}(t)&=& \ha_{\bfk i}(0) e^{-i\omega_\bfk t}-i g_\bfk
    \frac{\sin[(\omega_s-\omega_\bfk)t/2]}{(\omega_s-\omega_\bfk)t/2}
     \nonumber \\
    & &\times e^{-i(\omega_s+\omega_\bfk)t/2}  \hat{\sigma}_{\bfk4,1}.
\end{eqnarray}
Here, $g_\bfk$ is the atom-photon coupling constant,
$\omega_\bfk=c|\bfk|$ is the photon's frequency, and $\omega_s$ is
the resonance transition frequency. The
collective atomic operators are defined as
\begin{eqnarray}
    \hat{\sigma}_{2,\bfk3}=\hat{\sigma}_{\bfk3,2}^\dag=\sum_{\bfq,\bfQ} f(\bfk+\bfq-\bfQ) \hat{S}_{\bfQ 2 \bfq 3} \\
    \hat{\sigma}_{\bfk4,1}=\hat{\sigma}_{1,\bfk4}^\dag=\sum_{\bfq,\bfQ} f(\bfk+\bfq-\bfQ) \hat{S}_{\bfQ 4 \bfq 1}
\end{eqnarray}
where the ensemble operator $\hat{S}_{\mu \nu}$ and the structure function  $f(\bfk+\bfq-\bfQ)$ for the atomic sample are precisely defined in appendix \ref{Sec.APP.1}, Eq. (\ref{Smn}) and (\ref{fdf}), respectively. Inserting this result to the correlation function (\ref{eq.g2}) gives
\begin{eqnarray}\tt
\label{eq.g2.2}
    & & g^{(2)}(\bfk,-\bfk,\tau)=\frac{1}{T} \int_0^T dt \times \nl
       & &~~~
       \frac{\langle\hat{\sigma}_{\bfk3,2}(t) \hat{\sigma}_{1,\bfk4 }(t+\tau)
       \hat{\sigma}_{\bfk4,1 }(t+\tau)
       \hat{\sigma}_{2,\bfk3}(t)\rangle}{\langle\hat{\sigma}_{\bfk3,2}(t)\hat{\sigma}_{2,\bfk3}(t)\rangle
       \langle \hat{\sigma}_{1,\bfk4 }(t+\tau)
       \hat{\sigma}_{\bfk4,1 }(t+\tau)\rangle},
\end{eqnarray}
where any contribution from vacuum electromagnetic (EM) fluctuations, given by $\hat{a}_{\bfk s}(0)$ and $\hat{a}_{\bfk s}(0)$, vanishes when tracing over the EM vacuum.

As demonstrated in detail in Appendix \ref{Sec.APP.1}, because
our system is in a symmetric collective state throughout the dynamics, the products of coherence operators can be written as the products of number
operators. Applying this to equation (\ref{eq.g2.2}), we have
\begin{equation}
\label{Eq.BM.PPC.1}
    g^{(2)}\approx\frac{1}{T} \int_0^T
        dt\frac{\langle\hat{\sigma}_{\bfq3,2}(t) \hat{N}_{\bfq 4} (t+\tau)
      \hat{\sigma}_{2,\bfq3}(t)\rangle} {N_{\bfq 4} (t+\tau) N_2(t)(N_{\bfq 3} (t)+1)},
\end{equation}
where we have replaced the number operator $\hat{N}_1$ with its
meanfield $N_1$ and approximated $N_1+1\approx N_1$. This result
is a valid approximation as the $|1\rangle$ level is
macroscopically occupied, with $N_1\gg 1$. The discrete momentum
$\bfq$ is the nearest neighbor of $\bfk$, with which $|\bfk-\bfq|$ is minimized.

Following the standard procedure \cite{ScuZub97}, we write
\begin{eqnarray}
\label{Eq.BM.PPC.2}
    & &  N_{\bfq4} (t+\tau)=\sum_{j=1}^2 \chi_j(\tau) N_j(t)
    +\sum_{j=3}^4 \chi_j(\tau) N_{\bfq j}(t) \nl
     & &~~~~~~~~~~~+\left[\eta_1(\tau) \varrho_{12}(t)+
    \eta_2(\tau) \varrho_{\bfq 34}(t)+c.c\right],
\end{eqnarray}
where we have used the fact that the coherent terms $\langle
\hvr_{\bfq 32}\rangle$ and $\langle \hvr_{\bfq 41}\rangle$ are
zero throughout the dynamics, since neither pumping nor purely
spontaneous/superradiant decay processes will generate such
coherences. The coefficients $\chi_j(\tau)$ and $\eta_j(\tau)$ are
determined by studying the linear response of $N_{\bfq 4}(t+\tau)$
to small perturbations in each variable, via numerically solving
the rate equations (\ref{ndy})-(\ref{ndy1}). For instance, to
determine $\chi_3(\tau)$, we apply a small, instant perturbation
$\delta N_{\bfq 3}$ to the system dynamics, i.e., making $N_{\bfq
3}(t)\rightarrow N_{\bfq 3}(t)+\delta N_{\bfq 3}$. We then
calculate the resulting drift $\delta N_{\bfq 4}(t+\tau)$ from the
unperturbed value $N_{\bfq 4}(t+\tau)$, with which the coefficient
is determined as $\chi_3(\tau)=\frac{\delta N_{\bfq
4}(t+\tau)}{\delta N_{\bfq 3}(t)}$.

Following the quantum regression theorem \cite{Lax68}, in equation
(\ref{Eq.BM.PPC.1}) we write $\hat{N}_{\bfq4} (t+\tau)$ at time
$t+\tau$ in terms of operators at time $t$, using the result
(\ref{Eq.BM.PPC.2}). The resulting expression is then normalized and
factorized into products of the occupation numbers, giving
\begin{eqnarray}
\label{Eq.BM.PPC.3}
     g^{(2)} &=& 1+ \frac{1}{T} \int_0^T \!\!\!\!dt
    \left(\frac{\chi_3(\tau)-\chi_2(\tau)} { N_{\bfq 4}(t+\tau)}
    + \frac{\eta^\ast_1(\tau) \varrho_{12}(t)}
       {N_{2}(t) N_{\bfq 4}(t+\tau) }\right.\nn
     & +& \left.\frac{\eta^\ast_2(\tau) \varrho_{\bfq 43}(t)}
        {(N_{\bfq 3}(t)+1) N_{\bfq 4}(t+\tau)}\right).
\end{eqnarray}
As seen in figure \ref{fig2} (a) and (b), the atomic dynamics
undergoes quasi-steady-state evolution, with slow damping due to
atom losses. This allows us to approximate the populations and
coherences as constants for the time period of interests, while
effectively taking $T\rightarrow \infty$.

As seen in equation (\ref{Eq.BM.PPC.3}), a strong time correlation
between signal and idler photons requires $N_{\bfq 4}\ll 1$. Physically,
this is because the one-to-one correspondence between signal and
idler photons will be spoiled if there is more than one
`atom' in the same collective mode at the same time. This result
agrees with the established experimental criteria, where the driving
field must be weak such that most atoms remain in the initial ground
state \cite{BalBraKol05, KolDuBel06}. In this weak driving regime,
we find to a good approximation that
\begin{equation}
   g^{(2)}\approx 1+ \frac{1} {<N_{\bfq 4}>}\chi_3(\tau),
\end{equation}
with $<N_{\bfq 4}>$ being the time-averaging value of $N_{\bfq4}$.

For a weak coupling between $|3\rangle$ and $|4\rangle$ with
$\Omega_c\ll D\,\Gamma_4$, we find $\chi_3(\tau)\approx
\frac{|\Omega_c|^2}{D^2\Gamma^2_4}\ll 1$, which leads a time delay
$\frac{D\Gamma_4}{|\Omega_c|^2}$ between signal and idler photons.
A strong signal-idler correlation with $g^{(2)}\gg 1$ then
requires $N_{\bfq 4}$ to be very small. Since the pair generation
rate $R_I\sim N_{\bfq 4}$, this would ultimately limit the
achievable beam brightness of photon pairs. For a strong coupling
with $\Omega_c\gtrsim D\,\Gamma_4$, however, we find $\chi_3
(\tau)\approx \sin^2\left( \frac{1}{2}\Omega_{c}\tau\right)
\exp(-\frac{1}{2}D\,\Gamma_4\tau)\sim 1$. In this case, a
relatively larger $N_{\bfq 4}$ can give the same correlation, thus
elevating the obtainable beam brightness. The correlation function
in this case exhibits oscillatory and damped behaviors with sharp
peaks, associated with Rabi-oscillations between
$|3\rangle\leftrightarrow |4\rangle$ \cite{BalBraKol05}. The time
delay between signal and idler photons is roughly
$(D\,\Gamma_4)^{-1}$. We note in all regimes, the bandwidth of
signal and idler photons is the reciprocal of the delay time. In
figure \ref{fig2} (d), we plot the second-order correlation
functions for photon pairs propagating along $\pm\hat{z}$ and
$\pm\hat{x}$ directions. Both cases exhibit sharp peaks of widths
$\sim 0.05 \Gamma^{-1}$, indicating strong temporal correlation
which violates the standard Cauchy-Schwartz inequality by a factor
$\gtrsim 1000$.

\subsection{Polarization entanglement}
\label{Sec.BM.PE} 
\begin{figure}[htp]
\includegraphics[width=7.0cm]{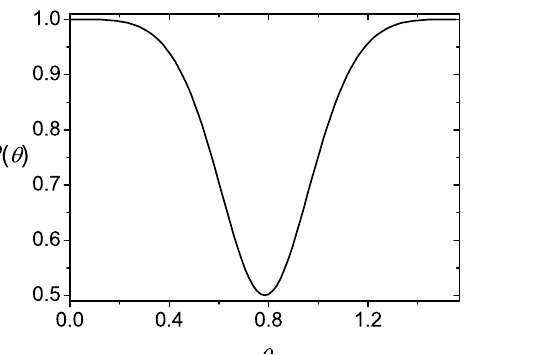}
    \caption{The probability $P(\theta)$ for paired
    photons to be in opposite polarizations as a function of emission angle
    $\theta$, calculated for the toy model.
 \label{fig3}}
\end{figure}

We now examine the polarization entanglement of
paired photons. We consider a pair of signal and idler photon
individually travelling along $\hat{k}$ and $-\hat{k}$ directions,
with
$\hat{k}(\theta,\phi)=\sin\theta\cos\phi\hat{x}+\sin\theta\cos\phi\hat{y}+\cos\theta\hat{z}$,
where $\hat{z}$ is along the $\bfK$ direction. For
$\hat{d}_{2,4}=\frac{1}{\sqrt{2}}(\hat{x}\pm i\hat{y})$, the
probabilities for signal photons to be left and right circularly
polarized along $\bfk$, denoted as $\hat{\epsilon}_L$
 and $\hat{\epsilon}_R$, are $\beta^{L}_{S}(\theta)=\left(1+\cot^4
\frac{\theta}{2}\right)^{-1}$ and
$\beta^R_{S}(\theta)=\beta^L_{S}(\pi-\theta)$, respectively.
Similarly, for the idler photons we have
 $\beta^R_{I}=\beta^L_S
(\theta)$, and $\beta^L_{I}=\beta^R_S (\theta)$. The probability for
photons to be in opposite circular polarizations along $\bfk$ (thus
in the same polarizations along each's propagating direction) is
then
\begin{equation}
    P(\theta)=\beta^L_s(\theta)\beta^R_I(\theta)+\beta^R_s(\theta)\beta^L_I(\theta)
    =\frac{1+\cot^8\frac{\theta}{2}}{\left(1+\cot^4
    \frac{\theta}{2}\right)^2}.
\end{equation}
As seen in Fig. \ref{fig3}, this is extremely flat around
$\theta=0,\pi$, where $P(\theta)\approx
1-\frac{1}{8}\textmd{mod}(\theta,\pi)^4\approx 1$, meaning that
photon pairs emitted over a wide range of $\theta$ will yield strong
polarization entanglement. Due to the temporal overlap of signal and
idler photons, each pair emitted within the strong correlation angle
is approximately in the Bell state of
$|\Psi^+\rangle=\frac{1}{\sqrt{2}}
(|\epsilon_R\epsilon_L\rangle+|\epsilon_L\epsilon_R\rangle)$. For
example, pairs with one photon emitted within $\theta<0.5$
(corresponding to $17\%$ of total emitted pairs) have an
entanglement fidelity $\ge99\%$.

\subsection{Critical Temperature} \label{Sec.BM.CD}

A biphoton source in an
atomic ensemble relies on collective enhancement (superradiance) to
increase the probability of correlation emission events, i.e. enforce `phase matching' for the light waves. A necessary
condition for collectivity is that the Doppler linewidth  is smaller than the photon's superradiance-broadened
linewidth, the latter of which is a factor $D$ greater
than the natural linewidth \cite{AreCou70,BonSchHaa71}. Physically,
this condition is required to suppress dephasing of the collective
atomic excitation during the superradiant emission process. The
present scheme is recoil-free in the sense that the driving and
coupling fields have net zero momenta, so that phase matching
requires only that the signal and idler photons have equal and
opposite momenta. However, an atom undergoing the biphoton emission
cycle does possess a nonzero recoil momentum while in the
intermediate states $|2\ra$, $|3\ra$ and $|4\ra$. Thus the system is
indeed subject to some Doppler broadening , as opposed to
conventional Doppler-free geometries where true two-photon pumping
creates an intermediate excited level with zero recoil momentum
\cite{Sal78}.

A useful alternative picture of Doppler broadening has emerged
from considering superradiant process in ultracold atomic gases
\cite{SteInoChi99,MooZobMey99}, based on the interplay between the
spatial coherence length and the recoil velocity. In this picture,
a thermal atom is viewed as a spatial `blob' of coherence, whose
size is given by the thermal coherence length
$\lambda_{coh}=\hbar/\sqrt{2mk_BT}$. When a single photon is {\it
collectively} absorbed, each atom is placed in a quantum
superposition of its initial state, and the excited state, which
is also a coherent blob, but one which is moving at the recoil
velocity relative to the initial blob. If collective excitation
lives too long,  then the initial and excited coherence blobs no
longer overlap in {\it space}, at which point it no longer matters
whether or not they overlap in momentum space. Overlap in both
momentum and position space is required in order for the
`which-atom' information to be erased. This sets a minimum
criterion for collective effects as,
$v_r\tau_{c}\ll\lambda_{coh}$, where $v_r=\hbar K/M$ is the recoil
velocity, and $\tau_c$ is lifetime of the collective excitation.
With $\Gamma_c=1/\tau_c$, this gives $\Gamma_c \gg \Gamma_d(T)$ as
the necessary condition for collective enhancement, where \beq
\Gamma_d(T)=K\sqrt{\frac{k_BT}{m}}, \label{Gammad}
\end{equation}
is the usual Doppler broadening contribution the linewidth.
We therefore see that Doppler broadening becomes significant only when
$\Gamma_d(T)$ is greater than or comparable with the linewidth of
signal and idler photons, both given by $D\, \Gamma_4$. The
threshold temperature of the system is then determined as
\begin{equation}
    T_c=\frac{m D^2 \Gamma^2_4}{k_B |\bfK|^2},
\end{equation}
above which collective effects disappear.
Taking typical $|\bfK|=10^7 \mathrm{m}^{-1}$, $D=100$ and
$m=10^{-25} \mathrm{kg}$, for $\Gamma_4=10^6 \mathrm{s}^{-1}$, $10^7 \mathrm{s}^{-1}$, and
$10^8 \mathrm{s}^{-1}$, we have $T_c=10$, $10^2$, and $10^4$ kelvin,
respectively. Due to the scaling of $T_c$ as $\Gamma_4^2$, we see that there could be a significant advantage to using
an atomic transition with a large natural linewidth.

This means the present butterfly scheme can be implemented with
room-temperature atomic clouds, excluding Cs which has $\Gamma\sim 10^6$, and so must be cooled below $10$K. We note that at intermediate
temperature, defined as $D\, \Gamma_4>\Gamma_d(T) \gg
\sqrt{\Delta^2+\Gamma_2^2}$, where $\Delta$ is the detuning of
$\Omega_d$ (which so far is taken to be zero), the effective
linewidth of level $|2\rangle$ is broadened from $\Gamma_2$ to
$\Gamma_d(T)$. For $\Delta=0$, this is the case when $T\gtrsim 1$
kelvin. The only consequences of this, however, are that the various
laser intensities and detunings would have to be adjusted
accordingly to maintain the condition (\ref{RIoverRS}), and the
sum-frequency linewidth would be similarly Doppler broadened. The pair correlations, however, will be unaffected. In contrast, for $T>T_c$ the pairing effect will completely disappear.

\section{Realistic Models}
\label{Sec.RM}

In the above section, we used a simplified toy model to illustrate
the physics of the butterfly scheme for ultra-bright photon pairs.
In this section, we provide two realizations of the butterfly
scheme, implemented with Ytterbium (section \ref{Sec.RM.SA}) and
Cesium atoms (section \ref{Sec.RM.CA}), respectively. The main
difference between toy and realistic models is that in the toy
model, we use effective Rabi frequencies to describe the
multi-photon pumping dynamics. In practice, this is an over-simplification of a
complex dynamical process, and
excludes the background Rayleigh/Raman scatterings associated with the intermediate states. To
study these features, in the following section, we shall expand the
toy dynamical model to account fully for the multi-photon nature of
the drive and control fields.

\subsection{Ytterbium Atoms}
\label{Sec.RM.SA}

\begin{figure}[htp]
\begin{center}
\includegraphics[width=8.0cm]{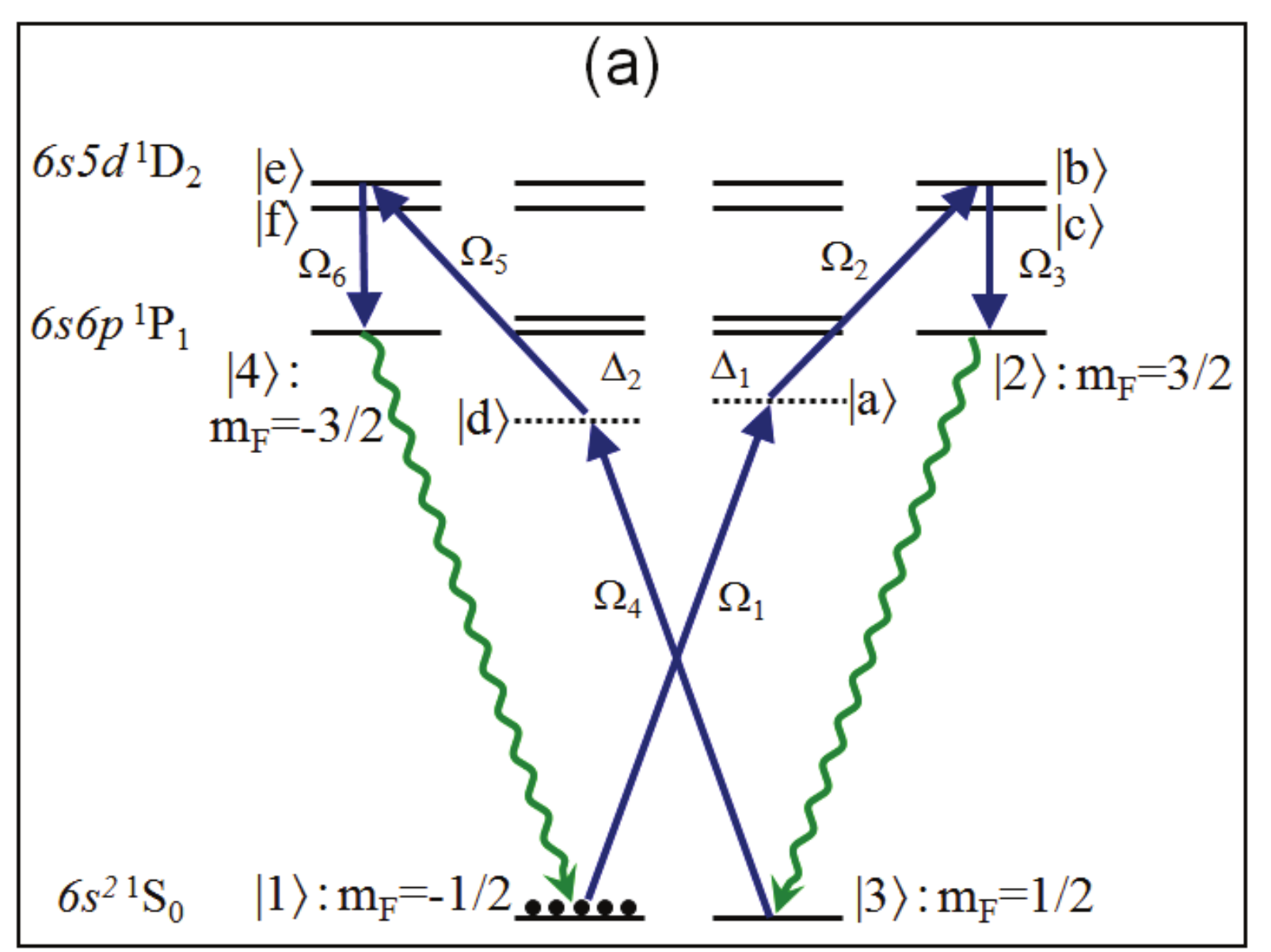}
\includegraphics[width=8.0cm]{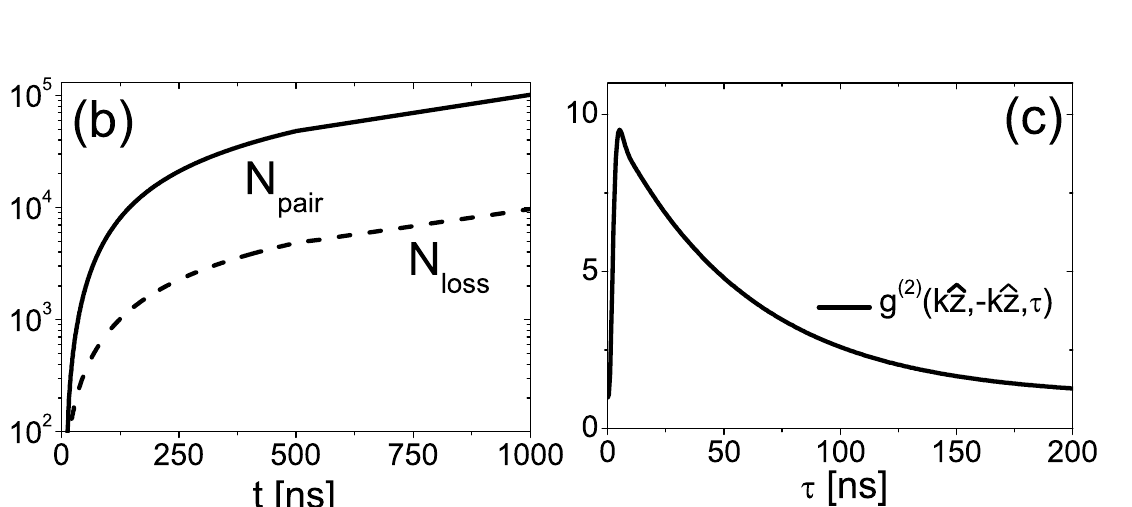}
    \caption{(Color online) A realistic butterfly scheme using
    $^{171}$Yb atoms. Similarly to figure \ref{fig4} for Cesium atoms,
    figure (a) shows the level diagram, (b) plots the time evolutions of
    $N_{pair}$ and $N_{loss}$, and (c) draws the time correlation function.
 \label{fig4}}
 \end{center}
\end{figure}

We now consider a realistic butterfly level scheme configured on
the $399$nm-line of the $6s^2$ $\mathrm{^1S_0}$$\leftrightarrow$
$6s6p$ $\mathrm{^1P_1}$ transition in $^{171}$Yb atoms, as shown
in figure \ref{fig4} (a). The atoms are prepared in the
$|1\rangle\equiv|F=1/2,m_F=-1/2\rangle $ state and then follow a
FWM cycle which deposits them in level $|3\rangle$. A second
independent FWM cycle then returns them to the initial $|1\rangle$
state. The driving FWM cycle consists of one violet laser and two
infrared lasers, with Rabi frequencies $\Omega_{1}$, $\Omega_{2}$
and $\Omega_{3}$. The coupling FWM cycle similarly contains one
violet and two infrared lasers, with Rabi frequencies of
$\Omega_{4}$, $\Omega_{5}$ and $\Omega_{6}$. The net momenta of
the driving lasers are equal and opposite with that of coupling
lasers. Signal photons are emitted as $|2\rangle\equiv
|F=3/2,m_F=3/2\rangle$ atoms spontaneously decay to the only
dipole-allowed state of $|3\rangle\equiv |F=1/2,m_F=1/2\rangle$,
and idler photons are generated as each atom in $|4\rangle\equiv
|F=3/2,m_F=-3/2\rangle$ decays collectively back to $|1\rangle$
with initial momentum $\hbar\bfk_0$. Both transitions have a
natural linewidth of $0.2$ GHz. Atoms which spontaneously decay to
other momentum modes of level $|1\rangle$ generate unpaired rogue
photons. Yet, they are still able to participate in the next-round
collective emission cycle. In contrast, atoms which spontaneously
decay from the intermediate pump levels to $|3\rangle$ are not
eligible for the collective emission. They nonetheless can be
repumped back to $|1\rangle$ in subsequent dynamics, during which
more rogue photons will be emitted. For simplicity, however, in
the present model we treat all unpaired-emission events as
permanent atom losses, thereby excluding any of the aforementioned
recycling or repumping processes. We note here the background
scatterings, involving spontaneous single-photon decay of
$m_F=-1/2, 1/2$ hyperfine states of the $6s6p$ level, are
efficiently suppressed by making $\Omega_1$ and $\Omega_4$
far-detuned from resonance. The intermediate $6s5d$ level has a
long lifetime of $6700$ns, compared to $5$ns for the $6s6p$ level,
so that the spontaneous decay is negligible.

In the present scheme, $\Omega_3$ and $\Omega_6$ correspond to
identical but counterpropagating lasers, and are thus
inter-changeable. In the driving process from $|1\rangle$ to
$|2\rangle$, an atom is equally likely to absorb a photon from
either of the two lasers. Then, during the coupling from
$|3\rangle$ to $|4\rangle$, it will preferably absorb a photon
from the other laser, following by collectively decay. This is
because for the competing process of absorbing a same-momentum
photon, the phase matching condition for collectivity is not
satisfied. Consequently, atoms driven by this channel only decay
spontaneously, the rate of which is a factor $1/D \ll 1$ smaller
than that of the collective, phase-matching channel.

The system's dynamics is solved by extending the rate equations
(\ref{ndy})-(\ref{ndy1}) to include intermediate pumping levels,
such as levels of $6s6p, m_F=-1/2, 1/2$ and $6s5d, m_F=-3/2,3/2$,
as well as important side transition channels, including atom loss
from $|1\rangle$ due to excitation by laser $\Omega_4$ and
$\Omega_5$. The resulting rate equations are obtained in a
straightforward manner very similar to the Cesium example derived
explicitly in the appendix \ref{Sec.APP.ERE}.

We consider a cold ($<1$ kelvin) spherical cloud of $10^6$ atoms
with diameter $L=26~\mathrm{\mu m}$, corresponding to an
optical depth of $D=20$. The six pumping lasers are chosen such
that the Rabi frequencies and detunings of transitions denoted in
Fig. \ref{fig4} yield values of (all in units of GHz)
$\Omega_1=5$, $\Omega_2=16$, $\Omega_3=2$, $\Omega_4=10$,
$\Omega_5=35$, $\Omega_6=2$, $\Delta_1=1000$, $\Delta_2=1200$.
With these parameter choices, the effective Rabi strength
resonantly coupling $|1\rangle$ to the level of $|F=3/2,
m_F=3/2\rangle$ is
$\Omega_{\mathrm{eff}}=\frac{\Omega_1\Omega_2}{2\Delta_1}=0.04$GHz,
obtained by adiabatically eliminating the intermediate $6s6p$
level using $\Omega_1, \Omega_2\ll \Delta_1$. The same lasers
couple $|1\rangle$ to another hyperfine level of
$|c\rangle\equiv|F=5/2, m_F=3/2\rangle$, but blue detuned from
resonance by $\Delta_{hf}=4$GHz. The effective Rabi frequency is
$\Omega_{\mathrm{eff}}/2\ll \Delta_{hf}$. This, aided by that the Rabi
frequency between $|c\rangle$ and $|2\rangle$ is
$2\Omega_3\not>\Delta_{hf}$, suppresses the pumping from
$|1\rangle$ to $|c\rangle$. This suppression necessarily provides
a non-vanishing coupling channel from $|1\rangle$ to $|2\rangle$.
Similarly, the coupling between $|3\rangle$ and
$|f\rangle\equiv|F=5/2,-3/2\rangle$, yielding an effective Rabi
frequency of $0.14$GHz$\ll \Delta_{hf}$, is suppressed, too. We
have numerically verified this analysis, where the population
ratio between level $|b\rangle\equiv|F=3/2, m_F=3/2\rangle$ and
$|c\rangle$ is found to be $4$, while the ratio between
$|e\rangle\equiv|F=3/2, m_F=-3/2\rangle$ and $|f\rangle$ is $14$.

With the above observations, we have solved the extended rate
equations numerically, with results shown in figure \ref{fig4} (b)
and (c). In figure (b), the production rate of photon pairs is about
$10^{11}$ per second, while the atom loss from initial momentum
modes is about $10^{10}$s$^{-1}$. The ratio of generated pairs
$N_{pair}$ to lost atoms $N_{loss}$ is then $10$. In figure (c), the
time correlation function exhibits sharp peaks, showing strong pair
correlation between signal and idler photons. The bandwidth of
signal and idler photons, measured by the reciprocal of correlation
time, is about $15$MHz, much smaller than the natural linewidth
$\Gamma_4=0.2$GHz. This is because the coupling between $|3\rangle$
and $|4\rangle$ is much weaker than the collective decay, so that
the system dynamics is in the overdamped regime.

The major atom loss is attributed to $|1\rangle$ atoms being
off-resonantly pumped to excited levels by $\Omega_1$ and
$\Omega_4$, and then decaying non-collectively by emitting rogue
photons. These rogue photons are different in frequency from
signal and idler photons by $\sim 1000$ GHz, which is much larger
than the bandwidth of single and idler photons ($\sim 0.02$GHz).
They can thus be easily filtered out from the photon-pair beams by
optical frequency-selectors, and therefore will not contribute to
the impurity of collected biphoton beams. By excluding such rogue
photons, the ratio of ``good'' paired photons to the remaining
rogue photons, which are nearly frequency-degenerate with the
photon pairs, turns out to be around $20$. For a twin beam of $N$
pairs, this results in a fluctuation of $\sqrt{N/20}$ in the
number difference between the two beams. In contrast, the
number-difference fluctuation for two uncorrelated coherent beams
of the same size is $\sqrt{2N}$. Hence, the generated twin beams
in the present setup yield a considerable number-squeezing factor
of $\sqrt{2N}/\sqrt{N/20}=2\sqrt{10}$.

\subsection{Cesium Atoms}
\label{Sec.RM.CA}

We now consider another implementation configured on the $852$-nm
line of $\mathrm{D}2$ transition in Cesium atoms. The $|1\rangle$,
$|2\rangle$, $|3\rangle$ states are corresponding to $|F=3,
m_F=3\rangle$, $|F=5, m_F=5\rangle$, $|F=4, m_F=4\rangle$ levels,
respectively. State $|4\rangle$ is consisted of degenerate $|F=4,
m_F=4\rangle$ and $|F=5, m_F=4\rangle$ hyperfine levels. Driving
from $|1\rangle$ to $|2\rangle$ is accomplished via a FWM process,
through detuned intermediate levels of $6\mathrm{P}_{3/2}$ and
$6\mathrm{D}_{5/2}$. The three driving lasers, with Rabi
frequencies $\Omega_1$, $\Omega_2$ and $\Omega_3$, are $\pi$,
$\sigma_+$ and $\sigma_-$ polarized, respectively. In order to
match the frequencies of signal and idler photons, the $\Omega_3$
laser is blue-detuned by the ground hyperfine splitting of $9.2$
GHz. We note since the lifetime of the $6D_{5/2}$ level ($\sim 1$
$\mu$s) is about $30$ times longer than that of $6p_{3/2}$ ($\sim
30$ ns), $\Omega_2$ can be tuned near- or on- resonant without
resulting in a faster atom loss than the photon-pair gain. The
coupling between $|3\rangle$ and $|4\rangle$ is provided by a
single resonant $\pi$-laser. Signal photons are emitted as
$|2\rangle$ atoms spontaneously decay to $|3\rangle$, while idler
photons are generated as $|4\rangle$ atoms collectively decay back
to $|1\rangle$. Thus different from the toy level scheme plotted
in Fig. \ref{fig1} and the Ytterbium scheme shown in Fig.
\ref{fig4}, the emitted photon pairs now have the same
polarizations. Collecting these pairs directly generates a highly
spin-squeezed beam in twin-Fock state, which is potentially useful
for precision interferometry measurement at Heisenberg-limited
sensitivity \cite{YurMcCKla86, HolBur93}, as well as on-demand
quantum teleportation among single-atom qubits \cite{HuaMoo08}.

The system's dynamics is solved by extending the rate equations
(\ref{ndy})-(\ref{ndy1}) to include all intermediate pumping
levels, such as levels $|a\rangle\equiv |6P_{3/2}, m_F=3\rangle$
and $|b\rangle\equiv |6D_{5/2}, m_F=4\rangle$, as shown in figure
\ref{fig4} (a). Each level is associated with a  corresponding
spontaneous decay rate to account for background Rayleigh and/or
Raman scattering of pumping lasers. Furthermore, there exist
undesired side transitions due to cross-driving by pumping lasers.
For example, atoms in $|1\rangle$ state are also driven by the
$\Omega_4$ laser to excited states, and $|3\rangle$ atoms are
additionally coupled to $|4\rangle$ by the $\Omega_1$ laser. These
transition are, however, far detuned from resonance. The net
effects are then atom losses from relevant levels, and can thus be
conveniently included in the rate-equation model by adding
appropriate loss rates to corresponding levels. In this way, the
multi-photon nature of driving and/or coupling is exactly treated via a
dynamical model, while the background scatterings involving side
transitions are calculated in a semi-exact manner, whose validity
is justified by large detunings. We present the extended set of
rate equations for the present Cesium scheme in appendix
\ref{Sec.APP.ERE}.

To examine the performance of the Cesium scheme, we consider a
spherical cloud of $10^6$ atoms with a diameter of $L=44
~\mathrm{\mu m}$, corresponding to $D=30$. The temperature of the
cloud is assumed to be well below $1$ kelvin, so that Doppler
broadening of level $|2\rangle$ is negligible compared to its
natural linewidth. We choose the pumping parameters as (all in
units of GHz): $\Omega_1=2$, $\Omega_2=100$, $\Omega_3=20$,
$\Omega_4=0.25$, $\Delta_1=300$, $\Delta_2=5$, $\Delta_3=9.2$. We
numerically solve the extended rate equations, with results shown
in figure \ref{fig4} (b) and (c). In figure (b), the production
rate of photon pairs is $1.7 \times 10^{11}$ per second, while the
atom loss rate is $0.2\times 10^{11} \mathrm{s}^{-1}$. The
photon-pair gain to atom loss ratio is then $\sim 9$. In figure
(c), the time correlation function exhibits sharp peaks,
indicating strong pair correlation between signal and idler
photons. The bandwidth of signal and idler photons, measured by
the reciprocal of correlation time, is about $100$MHz, which is
larger than the corresponding natural linewidth of $33$ MHz.

\begin{figure}[htp]
\begin{center}
\includegraphics[width=8.0cm]{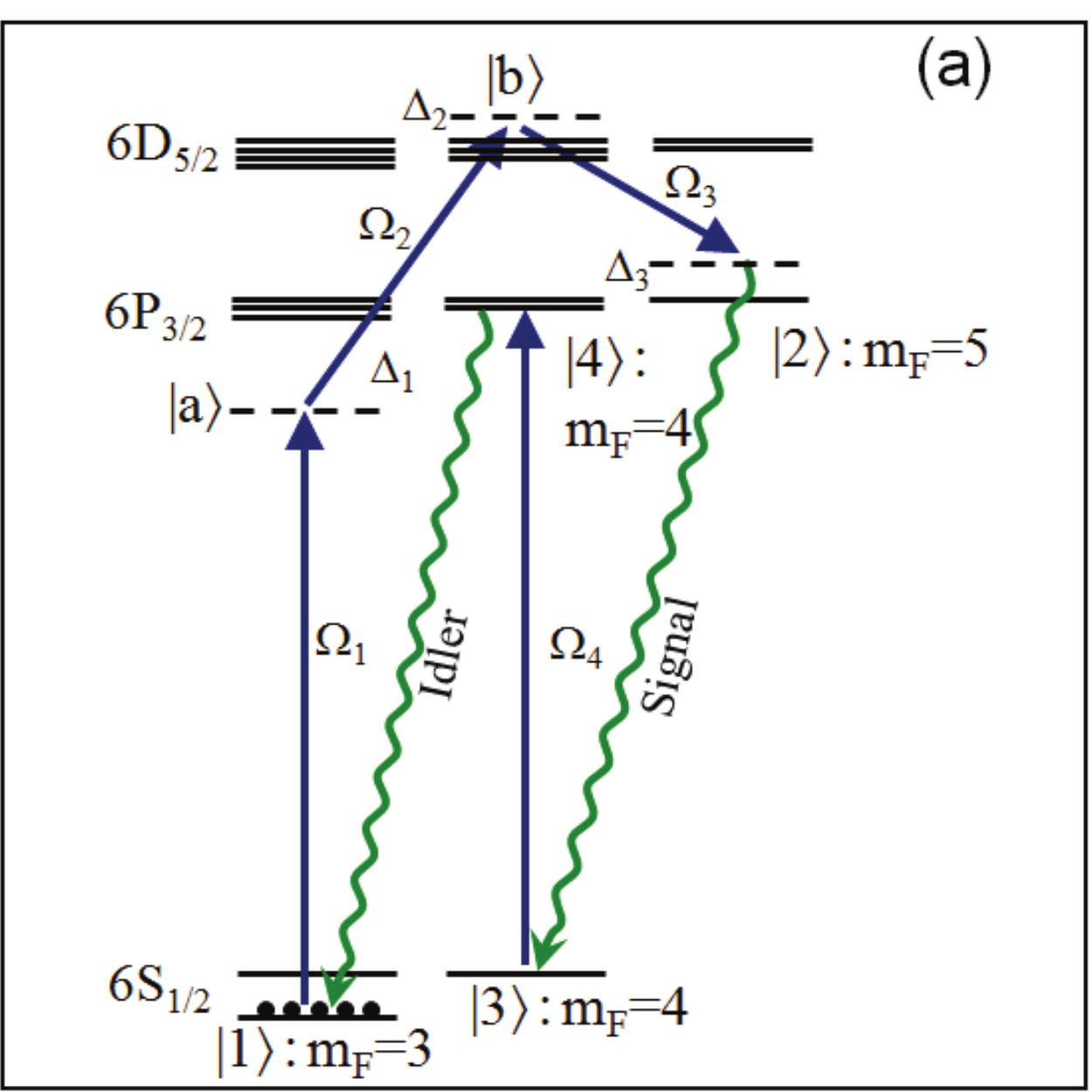}
\includegraphics[width=8.0cm]{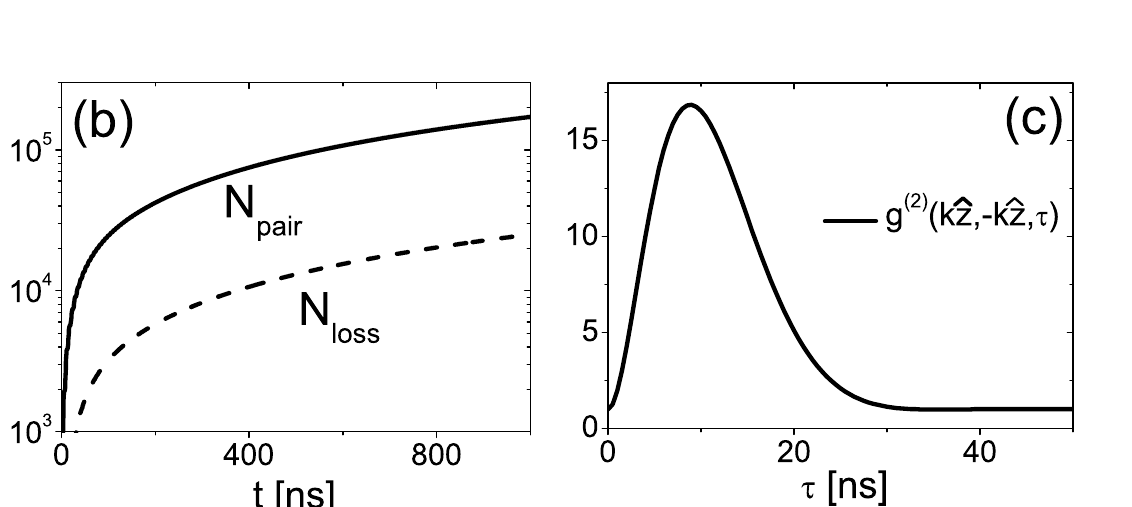}
    \caption{(Color online) A realistic butterfly scheme using
    Cesium atoms. Figure (a) shows the level diagram, (b) plots $N_{pair}$ and $N_{loss}$,
    and (c) shows the time correlation function
    for photon pairs emitted along $\pm \hat{z}$ directions
 \label{fig5}}
 \end{center}
\end{figure}

For the present setup, there are three dominating atom loss
channels. The first is via spontaneous relaxation of atoms in the
intermediate pumping level $|6D_{5/2},m_F=4\rangle$, imparting
emitting rogue photons at $917$nm wavelength. The second is via
background Rayleigh and/or Raman scattering of the $\Omega_1$ laser
by the populated $|1\rangle$ level, emitting photons at a frequency
$300$ GHz smaller than that of paired photons. The third channel is
via spontaneous emission of $|4\rangle$ atoms to $|3\rangle$ and the
ground $|F=4,m_F=3\rangle$ level, emitting rogue photons whose
frequency is $9.2$GHz smaller than the paired photons. All these
frequency-differences are much larger than the bandwidth of
generated photon pairs ($\sim 0.1$GHz), and can thus be filtered out
by optical frequency-selectors. By excluding the aforementioned
rogue photons, the ratio of ``good'' paired photons to the remaining
rogue photons in the collected biphoton beams turns out to $25$,
corresponding to a considerable squeezing factor of $5\sqrt{2}$.

\section{Reabsorption Analysis}
\label{Sec.RA} In any scheme using atomic ensembles for biphoton
sources, photon pairs are generated inside an optically-thick
atomic cloud. This raises the question of whether or not they are
able to propagate out of the cloud without reabsorption. In some
schemes \cite{VanEisAnd03,KolDuBel06,DuKolBel08}, suppression of
this reabsorption occurs naturally, due to the existence of an
inherent electromagnetically-induced transparency (EIT) window
\cite{HarFieIma90,ScuZub97,FleImaMar05}. For example, in Ref.
\cite{DuKolBel08}, where the generated photon pairs yield narrow,
sub-natural bandwidths due to overdamped coupling, the idler
photons are shown to propagate within the systems EIT window, thus
suppressing the reabsorption. The signal photons, on the other
hand, are off-resonance with respect to excitation of the only
populated $|1\rangle$ level, and thus in fact ``see'' an optically
thin medium.

In the Ytterbium scheme, similar EIT windows exist for both signal
and idler photons as well. For signal photons, the EIT window is
formed by the
$|1\ra\leftrightarrow|a\ra\leftrightarrow|^1\mathrm{D}_2,
m_F=1/2\ra$ $\Lambda$-structure. The signal photons weakly drive
the $|1\ra\leftrightarrow|a\ra$ transition, while the resonant
laser $\Omega_3$ strongly drives the
$|a\ra\leftrightarrow|^1\mathrm{D}_2, m_F=1/2\ra$ transition.
Since i) the $^1\mathrm{D}_2$ level (with a lifetime of $6700$ns)
is practically metastable compared to $^1\mathrm{P}_1$ (with a
lifetime of $5$ns), and ii) the couplings of $^1\mathrm{D}_2$ to
other levels are negligibly weak compared to $\Omega_3$, the
present $\Lambda$-structure is effectively mapped onto the
three-level EIT model \cite{VanEisAnd03,KolDuBel06,DuKolBel08}.
For idler photons, similar arguments apply, with the EIT window
formed by the $|1\ra\leftrightarrow| 4\ra\leftrightarrow|e\ra$
$\Lambda$-structure. These, together with the fact that the
generated photon pairs yield a sub-natural bandwidth, give rise to
similar EIT reabsorption-suppression effects demonstrated in
previous experiments \cite{KolDuBel06,DuKolBel08}, so that in the
present setup the majority of signal and idler photons will be
able to propagate out the atomic vapor.

\begin{figure}[htp]
\begin{center}
\includegraphics[width=8.0cm]{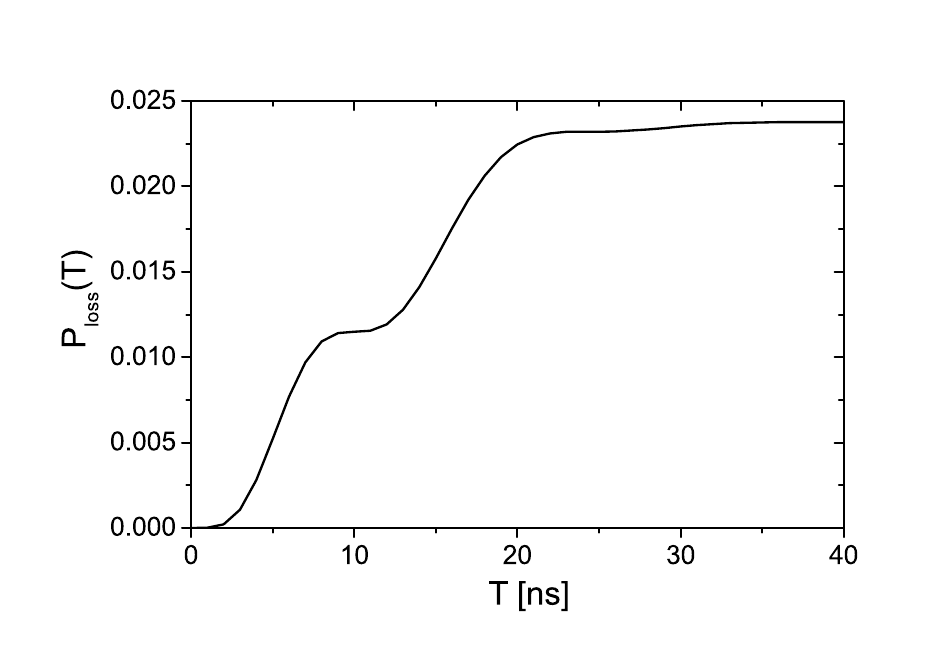}
    \caption{The loss probability $P_{loss}(T)$ as a function of time $T$ for
    both signal and idler photons in the cesium scheme.
 \label{fig6}}
 \end{center}
\end{figure}

In the Cesium scheme, the signal and idler photons, which are both
frequency and polarization degenerate, identically `see' a EIT
window formed by the
$|1\rangle\leftrightarrow|4\rangle\leftrightarrow|3\rangle$
$\Lambda$-structure, where the transition
$|4\rangle\leftrightarrow|3\rangle$ is resonantly driven by the
strong $\Omega_4$ laser, as in figure \ref{fig5} (a). However,
unlike the Ytterbium and previous schemes, here the bandwidth of
photon pairs is of the order of $1/10 \,\mathrm{ns}=0.1
\mathrm{GHz}$, as from figure \ref{fig5} (c). This bandwidth is
broadened from the natural linewidth $0.03 \mathrm{GHz}$. Thus, it
is unlikely that the EIT-suppression effect still applies, as the
bandwidth of photon pairs now matches the transparency window. On
the other hand, the pulse duration of the photon pairs is
$(D\Gamma_4)^{-1}$, whereas any scattering event occurs at a rate
not exceeding $\Gamma_4$. The probability for photon scattering is
then upper-bounded by $(D\Gamma_4)^{-1}\Gamma_4=1/D\ll 1$. To
verify this, in the following we use a simplified model to
simulate the absorption of a single-photon pulse passing through a
$\Lambda$-structure cloud. To account for the super-natural
bandwidth of $D\,\Gamma_4$, we approximate the single-photon Rabi
frequency $\Omega_y(t)$ with a ``toy'' envelope function of
\begin{equation}
   \Omega_y(t)=\Omega_0 e^{-D \Gamma_4 t/2} \sin(\Omega_4 t/2)
\end{equation}
where the coefficient
\begin{equation}
\Omega_0=\frac{8\Gamma_2
D}{\Omega_4}\sqrt{\frac{D^2\Gamma_4^2+\Omega_4^2}{N\pi}}
\end{equation}
is determined such that the underlying EM field yields the
single-photon energy $\hbar c|\bfK|$. The atomic dynamics of this
$\Lambda$ system is governed by a set of rate equations,
\begin{widetext}
\begin{eqnarray}
\label{Eq.RS.1}
       \ddt N_1 &=& \frac{i}{2}(\Omega_y \varrho_{41}-c.c)+\Gamma_{41}
    \bar{\beta} N_4(N_1+1), \\
    \ddt N_3 &=& \frac{i}{2}(\Omega_4 \varrho_{43}-c.c)+\Gamma_{43}
    \bar{\beta} N_4(N_3+1), \\
     \ddt N_4&=&-\frac{i}{2}(\Omega_4 \varrho_{43}+\Omega_y
    \varrho_{41}-c.c) -\Big(\Gamma_4+\bar{\beta}(\Gamma_{41} N_1+\Gamma_{43} N_3)\Big) N_4, \\
    \ddt
    \varrho_{41}&=&-i\frac{\Omega_{y}}{2}(N_4-N_1)+i\frac{\Omega_4}{2}\varrho_{31}
    -\left(\frac{\Gamma_4}{2}\right.
     \left.-\frac{\Gamma_{41}}{2} \bar{\beta}(N_4-N_1)
    +\frac{\Gamma_{43}}{2} \bar{\beta}N_3\right)\varrho_{41}, \\
     \ddt
    \varrho_{43}&=&-i\frac{\Omega_{4}}{2}(N_4-N_3)+i\frac{\Omega_y}{2}\varrho^\ast_{31}
    -\left(\frac{\Gamma_4}{2}\right.
    \left.-\frac{\Gamma_{43}}{2} \bar{\beta}(N_4-N_3)
    +\frac{\Gamma_{41}}{2} \bar{\beta}N_1\right)\varrho_{43}, \\
    \label{Eq.RS.2}
    \ddt\varrho_{31}&=&-i\frac{\Omega_y}{2}\varrho^\ast_{43}+i\frac{\Omega_4}{2}\varrho_{41}
     +\left(\frac{\Gamma_{43}+\Gamma_{41}}{2}
    \bar{\beta}N_4\right)\varrho_{31},
\end{eqnarray}
\end{widetext}
obtained in a similar manner with Eq. (\ref{ndy})-(\ref{ndy1}).
Here, $N_1$, $N_3$ and $N_4$ are populations at level $|1\rangle$,
$|3\rangle$ and $|4\rangle$. $\varrho_{ij}$ ($i,j=1,3,4$) is the
coherence between level $|i\rangle$ and $|j\rangle$. $\Gamma_{ij}$
is the spontaneous emission rate from $|i\rangle$ to $|j\rangle$
and $\Gamma_i=\sum_j \Gamma_{ij}$ is the natural linewidth of
level $|i\rangle$. Starting with all atoms in the $|1\rangle$
state, the above equations of motion are solved numerically.
Because photon losses correspond to scattering atoms out of the
$\Lambda$ system, the loss probability $P_{loss}$ of the photon
passing through the medium is given by the atomic population
reduction,
\begin{equation}
   P_{loss}(T)=N_1\Big|_{t=0}-(N_1+N_2+N_3)\Big|_{t=T}.
\end{equation}
A plot of $P_{loss}(T)$ is shown in figure \ref{fig6}, for
parameters used in section \ref{Sec.RM.CA}. We find the ultimate
loss probability $P_{loss}(\infty)\approx 0.02$ for both signal
and idler photons. This result is consistent with the upper-limit
of $1/D\approx 0.03$ as from the time-scale argument. In practice,
this photon loss due to absorption will lead to a $2\%$
degradation of the pair correlation.

\section{Conclusion and Discussions}
\label{Sec.Conc}

In this paper we have proposed an omnidirectional biphoton source based
on collective emission in an atomic vapor. This is accomplished by
employing multi-photon excitation and `Doppler-free' pumping. Our
scheme benefits from the elimination of background Rayleigh
scattering via dipole-selection rules, and the fact that
phase-matching can be fullfilled in the full $4\pi$ solid angle. We
have demonstrated an achievable photon-pair brightness of several
orders of magnitude greater than the best reported results. Our
scheme has the unique feature of strongly suppressed atomic
re-thermalization, thus allowing implementations in `hot' vapors
as well as ultracold samples such as BEC's, performed
nondestructively. This may lead to nondestructive \emph{in situ}
imaging of condensates, with an anticipated atom-counting precision below the
standard quantum limit. This is because potentially, more photons
than atoms can be generated before destroying the condensate.
Furthermore, we found strong time and polarization correlation
between signal and idler photons. We also provided two realistic
implementation of the present scheme, using Cesium and Ytterbium
atoms. In both schemes, the generated photon pairs can propagate
through the optically thick cloud without being reabsorbed.

In the present scheme, the impurity of collected biphoton beams
are attributed to i) non-collective emissions from level
$|4\rangle$, and ii) background Raman/Rayleigh scattering of
pumping lasers by the $|1\rangle$ atoms. The branch ratio of
collective to non-collective emission is $D$, i.e., the optical
depth of the atomic sample. One then might seek to improve the
beam purity by employing samples of higher optical depth. The
difficulty is that this requires a stronger coupling between
$|3\rangle$ and $|4\rangle$, in order to maintain a strong
time correlation between signal and idler photons. In both Yb and
Cs schemes, this will unavoidably lead to a stronger background
scattering, adding to the beam impurity.

Hence, an optimal
optical depth is in fact obtained by balancing the two impurity sources of  non-collective emission from level $|4\rangle$ and Raman/Rayleigh scattering of pumping lasers by the $|1\rangle$ atoms.
To strongly drive the $|3\rangle\leftrightarrow|4\rangle$ transition while avoiding significant background scattering, it is clearly necessary
to seek appropriate level structures where the coupling
is resonant for the $|3\rangle\leftrightarrow
|4\rangle$ transition, but far detuned from exciting
$|1\rangle$ atoms. An example is seen in the Cesium scheme in
section \ref{Sec.RM.CA}, where the coupling laser $\Omega_4$,
while resonant with the $|3\rangle\leftrightarrow |4\rangle$
transition, is red-detuned from the $|1\rangle$
excitation due to ground hyperfine splitting. For further improvements, one
may seek level scheme configured on the fine
structure to exploit the extremely large level splitting ($\gtrsim$ THz). Feasible atomic species include those having no  ground state hyperfine splittings, such as $^{120}$Sn and $^{28}$Si. For these atoms, we have found butterfly level structures similar to the cesium scheme shown in figure \ref{fig5} (a), but
configured on fine structure levels, where the spacing between
level $|1\rangle$ and $|3\rangle$ is now $10^4\sim 10^5$ GHz,
instead of $9.2$ GHz as for Cesium. These schemes are expected to have very low background scattering. The disadvantage, however, is that because the omnidirectional phase-matching condition relies
on producing (nearly) identical-wavelength signal and idler
photons, the $|2\rangle$ level must be blue-detuned by an amount
of ground fine splitting of $10^4\sim 10^5$ GHz, which would then
require ultra-intense driving lasers, making it difficult implement a cw scheme experimentally. Future work will therefore study the feasibility and applicability of a pulsed driving scheme.

If desired, level structures similar to the Cesium scheme in figure
\ref{fig5} (a) can be applied to other Alkaline atoms, such as Rb
and Na. However, they have smaller ground hyperfine splittings than
Cs, and will thus be subject to stronger background scattering. To overcome
this difficulty, one possibility would be to increase the
splitting by applying a static magnetic field, making use of the fact that
level $|1\rangle$ and $|3\rangle$ are exposed to different Zeeman
shifts. For a rough estimation, increasing the splitting to $10$ GHz only requires a magnetic field of $\sim 0.1$ T, suggesting experimental
feasibility.

Lastly, we note that as the atom loss fraction becomes significant,
the system will enter a more complicated regime, where the system
tries to equilibrate, potentially resulting in macroscopic occupation of all
ground hyperfine sub-levels, and presumably strongly diminished pair
correlations. If this is the case, only a small number of photon
pairs will be generated, but in a very short bright initial burst. Thus,
the butterfly system may be an excellent source for generating
highly number-difference squeezed twin pulses for quantum
interferometry. Further studies of the butterfly system in this
equilibrium regime remain as a future task.

Acknowledgement: we thank A. Leanhardt for helpful discussions. This
work is supported in part by National Science Foundation Grant No.
PHY0653373.

\begin{appendix}

\section{Derivation of Rate equations}
\label{Sec.APP.1}
In this appendix, we describe how one can derive the rate equations
(\ref{ndy})-(\ref{ndy1}) from first principles, based on a Markovian treatment of the optical field.
We begin by deriving a coarse-grained version of the Heisenberg equation of motion for an arbitrary operator, $\hA$. This approach is equivalent to the standard master equation approach, but eliminates the density operator, as it is an unnecessary intermediate step when the goal is to derive equations for expectation values of observables.

The standard quantum time propagator is given by $\hU(t)=e^{-i\hH t}$. Defining in the usual way, the Heisenberg picture operator, $\hA(t,t_0)=\hUd(t-t_0)\hA\hU(t-t_0)$, it the follows that $\hA(t+\tau,t_0)=\hUd(t+\tau-t_0)\hA\hU(t+\tau-t_0)$, so that
\beqa
\ddt\hA(t,t_0)&=&\lim_{\tau\to0}\frac{1}{\tau}\left(\hUd(t+\tau-t_0)\hA\hU(t+\tau-t_0)\right.\nn
&-&\left.\hUd(t-t_0)\hA\hU(t-t_0)\right)
\eeqa
We can then set $t_0=t$ to arrive at
\beq
\label{ddtA}
\ddt\hA=\frac{\hUd(\tau)\hA\hU(\tau)-\hA}{\tau}
\eeq
where $\hA$ is now the Schr\"odinger picture operator, and $\tau$ must be chosen sufficiently small. Taking the expectation value of this equation with respect to the state of the system at time $t$, will then yield the time-derivative of the expectation value at time $t$. The Markov approximation then consists of letting $\tau$ be small compared to the system evolution timescale, but long compared to the dephasing time of the reservoir.

Using this approach, we now derive a generic equation of motion for the toy level scheme of Sec. \ref{Sec.BM.TLS}. We start from a generic system-reservoir model,
\beq
\label{Hfull}
\hH=\hH_s+\hH_r+\hV_{sr},
\eeq
 where the reservoir consists of a bath of bosonic field modes, governed  by a  Hamiltonian of the form
\beq
\label{Hr}
\hH_r=\sum_\bfk(\omega_\bfk-\omega_s)\had_\bfk\ha_\bfk,
\eeq
where $\ha_\bfk$ annihilates a bath particle with wave-vector, $\bfk$, and frequency, $\omega_\bfk$, and $\omega_s$ is the resonance
frequency for the system-reservoir interaction. We assume an interaction operator of the form
\beq
\hV_{sr}=\sum_\bfk g_\bfk \had_\bfk\hc_\bfk+h.c.,
\eeq
where $\hc_\bfk$ is an unspecified system operator.

Now let us evaluate (\ref{ddtA}) for the case $\hA\to\hS$, where $\hS$ is a system operator only. To second-order in $\hV_{sr}$, this gives
\begin{widetext}
\beqa
\ddt\hS&=&\frac{1}{\tau}\left[\hUd_0(\tau)\hS\hU_0(\tau)-\hS\right]
+\frac{1}{\tau}\int_0^\tau dt_2\int_0^\tau dt_1 \hUd_0(t_1)\hV_{sr}(t_1)\hUd_0(\tau-t_1)\hS\hU_0(\tau-t_2)\hV_{sr}(t_2)\hU_0(t_2)\nn
&-&\frac{1}{\tau}\int_0^\tau dt_2\int_0^{t_2}dt_1\hU_0(\tau-t_2)\hS\hU_0(\tau-t_2)\hV_{sr}(t_2)\hU_0(t_2-t_1)\hV_{sr}(t_1)\hU_0(t_1)\nn
&-&\frac{1}{\tau}\int_0^\tau dt_2\int_0^{t_2}dt_1\hUd_0(t_1)\hV_{sr}(t_1)\hUd_0(t_2-t_1)\hV_{sr}(t_2)\hUd_0(\tau-t_2)\hS\hU_0(\tau)
\eeqa
\end{widetext}
where $U_0(t)=e^{-i(H_s+H_r)t}$, and we have neglected the first-order terms as they will vanish when we trace over the reservoir degrees of freedom. For the Markov approximation, we first take
\beq
U_0(\tau)\approx (1-i\tau \hH_s)e^{-i\hH_r\tau},
\eeq
and then perform the reservoir trace. The time integrals are then handled via,
\beq
\int_0^{t_2}dt_1e^{-i(\omega_\bfk-\omega_a)(t_1-t_2)}\approx\pi\delta(\omega_\bfk-\omega_a),
\eeq
which gives
\beqa
\ddt \hS&=&\sum_{\bfk}\pi\delta(\omega_\bfk-\omega_s)|g_\bfk|^2\left[\left[\hcd_\bfk,\hS\right]\hc_\bfk+\hcd_\bfk\left[\hS,\hc_\bfk\right]\right]\nn
&+&i\left[\hH_s,\hS\right]
\label{dSdt}
\eeqa
If one desires, the equation of motion for the system density operator, $\hrho_s$, can be obtained as a special case of (\ref{dSdt}), with the substitution $\ddt\to-\ddt$, as the state of the system must be evolved backwards in time to obtain the state in the Heisenberg picture.

For the toy model depicted in Fig. \ref{fig1}, the system Hamiltonian is
\beqa
\label{Hs}
\hH_s&=&\sumj\left[\Odt\left(\eKrj|2\ra\la 1|\o{j}+\emKrj|1\ra\la 2|\o{j}\right)\right.\nn
&+&\left.\Oct\left(\emKrj|4\ra\la3|\o{j}+\eKrj|3\ra\la 4|\o{j}\right)\right],
\eeqa
where $\hat{\bfr}_j$ is the position operator of the $j^{th}$ atom, and the state $|m\ra\o{j}$ indicates that the $j^{th}$ atom is in internal state $|m\ra$. The system Hamiltonian (\ref{Hs}) is given for a frame rotating at the system resonance frequency, $\omega_s$, corresponding to the frequency of the  $|4\ra\to|1\ra$ and $|2\ra\to|3\ra$ transitions. The system-reservoir interaction is described by the system operators
\beq
\label{ck}
\hc_\bfk:=\sumj \emkrj\left(|1\ra\la 4|\o{j}+|3\ra\la2|\o{j}\right).
\eeq

For the initial state of the system, we assume that each atom is an internal state $|1\ra$, and occupies a single box eigenstate, $\bfq$, so that
\beq
\label{psii}
|\psi_i\ra=\prodj |\bfq_j\ra\o{j}\otimes|1\ra\o{j}.
\eeq
 The initial momentum of the $j^{th}$ atom, $\bfq_j$ is to be chosen at random from the Boltzman distribution. For large enough $N$, this will reproduce the results of thermal averaging in a single realization. We then introduce the set of states
\beqa
\label{uq1j}
|u_{\bfq1}\ra\o{j}&:=&|\bfq_j+\bfq\ra\o{j}\otimes|1\ra\o{j}\\
\label{uq2j}
|u_{\bfq2}\ra\o{j}&:=&|\bfq_j+\bfK+\bfq\ra\o{j}\otimes|2\ra\o{j},\\
\label{uq3j}
|u_{\bfq,3}\ra\o{j}&:=&|\bfq_j+\bfK+\bfq\ra\o{j}\otimes|3\ra\o{j},\\
\label{uq4j}
|u_{\bfq4}\ra\o{j}&:=&|\bfq_j+\bfq\ra\o{j}\otimes|4\ra\o{j},
\eeqa
so that for the $j^{th}$ atom, $|u_{{\bf 0}1}\ra\o{j}\ra$ and  $|u_{{\bf 0}2}\ra\o{j}$  are the initial state and the state after interacting with the drive lasers, respectively.  The states $|u_{\bfq3}\ra\o{j}$ and $|u_{\bfq4}\ra\o{j}$, then correspond to the state after emission of a signal photon with momentum $\bfk\approx -\bfq$, and the state after then absorbing a sequence of photons from the coupling lasers, respectively. Emission of an idler photon with $\bfk\approx \bfq$, will then return the atom to the $|u_{\bfz1}\ra\o{j}$; whereas emission of a rogue photon will transform the state into $|u_{\bfq1}\ra$ with $\bfq\neq 0$.

With respect to the states (\ref{uq1j})-(\ref{uq4j}), the system Hamiltonian becomes
\beq
\label{Hsu}
\hH_s=\sum_\bfq\left[\Odt\left(\hS_{\bfq1\bfq2}+\hS_{\bfq1\bfq2}^\dag\right)+\Oct\left(\hS_{\bfq3\bfq4}+\hS_{\bfq3\bfq4}^\dag\right)\right],
\eeq
where we have introduced the generic atomic transition operators,
\beq
\label{Smn}
\hS_{\mu\nu}=\hS_{\nu\mu}^\dag:=\sumj |u_\mu\ra\la u_\nu|\o{j},
\eeq
where $\mu,\nu\in\{\{\bfq1\},\{\bfq2\},\{\bfq3\},\{\bfq4\}\}$ are  composite indices. Similarly, the interaction operators take the form
\beq
\label{cku}
\hc_\bfk=\sum_{\bfq,\bfQ}f(\bfk+\bfq-\bfQ)\left(\hS_{\bfq1\bfQ4}+\hS_{\bfq3\bfQ2} \right),
\eeq
where
\beqa
\label{fdf}
f(\bfk{+}\bfq{-}\bfQ)&=&\la \bfq|\emkrj|\bfQ\ra\nn
&=&\frac{1}{V}\int_Vd^3r\, e^{-i(\bfk+\bfq-\bfQ)\cdot\bfr}
\eeqa
is the static structure function of the sample.

The operators we are interested in are all one-body operators of the form (\ref{Smn}). In order to evaluate (\ref{dSdt}) for these operators, we will need to evaluate commutators of the form,
\beq
\left[\hS_{\mu\nu},\hS_{\alpha,\beta}\right]=\delta_{\nu,\alpha}\hS_{\mu\beta}-\delta_{\mu,\beta}\hS_{\alpha\nu}.
\label{commSS}
\eeq
Taking the expectation value of the equation of motion (\ref{dSdt}) will then require us to compute the expectation value of the bilinear operator $\la \hS_{\mu\nu}\hS_{\alpha\beta}\ra$. Our strategy for dealing with these terms will make use of the underlying exchange symmetry which leads to the emergence of collectivity in the emission properties of the sample.

We note that the initial state (\ref{psii}) is not symmetric under particle label exchange, due to the dependence of the state $|u_{\bfz1}\ra\o{j}$ on $\bfq_j$. However, as the set of states $\{|u_{\bfq1}\ra\o{j},|u_{\bfq2}\ra\o{j},\{|u_{\bfq3}\ra\o{j}\},\{|u_{\bfq4}\ra\o{j}\}\}$ forms a complete basis, all expectation values of operators of the the form (\ref{Smn}) will depend only on inner-products of the form
$\la u_\mu|u_\nu\ra\o{j}$, which are in fact independent of the $\bfq_j$'s. For example, with $m,M\in\{1,2,3,4\}$, we have
\beqa
\la u_{\bfq m}| u_{\bfQ M}\ra\o{j}&=&\la m|\o{j}\otimes\la \bfq_j{+}\bfq|\o{j}| M\ra\o{j}\otimes|\bfq_j{+}\bfQ\ra\o{j}\nn
&=&\delta_{m,M}\,f(\bfq-\bfQ)
\eeqa
which does not depend on $\bfq_j$. This means that we can make any choice we like for the set of $\bfq_j$'s, without affecting the rate equation dynamics. We note that this occurs in part because we have neglected the kinetic energy of the atomic center-of-mass motion in our system Hamiltonian (\ref{Hs}), an approximation valid when the lifetime of a collective excitation is short compared to $1/(\omega_R+\Gamma_d(T))$, where $\omega_R=\hbar K^2/2M$ is the recoil frequency, and $\Gamma_d$ is the Doppler line-width (\ref{Gammad}).

One option is then to set all the $\bfq_j$'s to zero, in which case our initial state becomes explicitly symmetric under exchange of particle label exchange. Equivalently, one can recognize that the $\bfq_j$'s are redundant with the particle labels, so that exchange of the $\bfq_j's$ is part of the underlying symmetry. As the Hamiltonian is also symmetric under this form of particle exchange, it follows that the state of the full system will remain symmetric as it evolves in time. The full state of the system + reservoir can then be written as
\beqa
\label{symm}
|\psi_{SR}(t)\ra=\sum_{\mu_1,\ldots,\mu_N}\sum_r c(\mu_1,\ldots,\mu_N;r)|u_{\mu_1}\ra\o{1}\otimes|u_{\mu_2}\ra\o{2}\nn
\otimes|u_{\mu_3}\ra\o{3}\ldots\otimes|u_{\mu_N}\ra\o{N}\otimes|r\ra\o{R},\ \ \ \ \
\eeqa
where $r$ is a composite index which sums over all states of the reservoir, and the state $|r\ra\o{R}$ lives in the reservoir Hilbert space. Particle exchange symmetry then requires that $c(\mu_1,\ldots,\mu_N;r)$ be invariant under exchange of any two $\mu_j$'s.

With the symmetric state (\ref{symm}), we can now evaluate the expectation value of the product of two operators of the form (\ref{Smn}),
\beq
\la\hS_{\mu\nu}\hS_{\alpha\beta}\ra=\delta_{\nu\alpha}\la\hS_{\mu\beta}\ra
+\sum_{{\ss j,J=1}\atop{\ss J\neq j}}^N\la\hS_{\mu\nu}\o{j}\hS_{\alpha\beta}\o{J}\ra,
\label{Smnab}
\eeq
where
\beq
\hS_{\mu\nu}\o{j}:= |u_\mu\ra\la u_\nu|\o{j},
\eeq
and with $\mu=\bfq m$ and $\nu=\bfQ M$, we have introduced
\beq
\delta_{\mu\nu}=\delta_{\bfq,\bfQ}\delta_{m,M}.
\eeq
To illustrate an important consequence of exchange symmetry on the bilinear terms in (\ref{Smnab}), we consider first the $j{=}1$, $J{=}2$ case,
\begin{widetext}
\beqa
\label{Smban}
\la \hS_{\mu\nu}\o{1}\hS_{\alpha\beta}\o{2}\ra&=&\la\psi(t)| \hS_{\mu\nu}\o{1}\hS_{\alpha\beta}\o{2}|\psi(t)\ra
\nn&=&
\sum_{{\ss  \mu_1,\ldots,\mu_N}\atop{\ss \nu_1,\ldots,\nu_N}}\sum_{r}c\str(\mu_1,\ldots,\mu_N;r)
c(\nu_1,\ldots,\nu_N;r)
\la u_{\mu_1}|u_\mu\ra\la u_\nu| u_{\nu_1}\ra
\la u_{\mu_2}|u_\alpha\ra\la u_\beta|u_{\nu_2}\ra \la u_{\mu_3}|u_{\nu_3}\ra\ldots\la u_{\mu_N}|u_{\nu_N}\ra\nn
&=&\sum_{\mu_3,\ldots,\mu_N}\sum_rc\str(\mu,\alpha,\mu_3,\ldots,\mu_N;r)c(\nu,\beta,\mu_3,\ldots,\mu_N;r)\nn
&=&\sum_{\mu_3,\ldots,\mu_N}\sum_rc\str(\mu,\alpha,\mu_3,\ldots,\mu_N;r)c(\beta,\nu,\mu_3,\ldots,\mu_N;r)\nn
&=&\la \hS_{\mu\beta}\o{1}\hS_{\alpha\nu}\o{2}\ra,
\eeqa
\end{widetext}
where we have  used the relation $\la u_\mu|u_\nu\ra=\delta_{\mu,\nu}$ to eliminate these inner products.
Because the state is symmetric under exchange of particle labels,  Eq. (\ref{Smban}) can be generalized to
$\la \hS_{\mu\nu}\o{j}\hS_{\alpha\beta}\o{J}\ra=\la \hS_{\mu\beta}\o{j}\hS_{\alpha\nu}\o{J}\ra$, for $j\neq J$, which gives us
\beq
\label{Smban2}
\la\hS_{\mu\nu}\hS_{\alpha\beta}\ra=\delta_{\nu,\alpha}\la\hS_{\mu\beta}\ra
+\sum_{{\ss j,J=1}\atop{\ss J\neq j}}^N\la\hS_{\mu\beta}\o{j}\hS_{\alpha\nu}\o{J}\ra,
\eeq
as an equivalent alternative to (\ref{Smnab}). This result will allow us to express the equations of motion for number operators in terms of products of number operators, as opposed to products of coherence operators.

The equation of motion for the expectation value of a system operator of the form (\ref{Smn}) is then
\beqa
\label{dSmndt}
\ddt\la\hS_{\mu\nu}\ra&=&\sumk\pi\delta(\omega_\bfk{+}\omega_s)|g_\bfk|^2
\left(\la\left[\hc_\bfk^\dag,\hS_{\mu\nu}\right]\hc_\bfk\ra\right.\nn
&+&\left.\la\hc_\bfk^\dag\left[\hS_{\mu\nu},\hc_\bfk\right]\ra\right)+
i\la\left[\hH_s,\hS_{\mu\nu}\right]\ra.\nn
\eeqa
We can evaluate the commutators via Eqs. (\ref{Hsu}), (\ref{cku}) and (\ref{commSS}), resulting in
\beqa
\label{ckdSmn}
\left[\hc_\bfk^\dag,\hS_{\mu\nu}\right]
&=&\sum_{\bfq,\bfQ}f\str(\bfk{+}\bfq{-}\bfQ)\left(\delta_{\mu,\bfq1}\hS_{\bfQ4\nu}-\delta_{\nu,\bfQ4}\hS_{\mu\bfq1}\right.\nn
&&+\left.\delta_{\mu,\bfQ2}\hS_{\bfq3\nu}-\delta_{\nu,\bfq3}\hS_{\mu\bfQ2}\right),
\eeqa
\beqa
\label{Smnck}
\left[\hS_{\mu\nu},\hc_\bfk\right]
&=&\sum_{\bfq,\bfQ}f(\bfk{+}\bfq{-}\bfQ)\left(\delta_{\nu,\bfq1}\hS_{\mu\bfQ4}-\delta_{\mu,\bfQ4}\hS_{\bfq1\nu}\right.\nn
& &+\left. \delta_{\nu,\bfQ2}\hS_{\mu\bfq3}-\delta_{\mu\bfq3}\hS_{\bfQ2\nu}\right).
\eeqa
and
\beqa
\label{HsSmn}
\left[\hH_s,\hS_{\mu\nu}\right]=\Odt\sumq\left(\delta_{\mu\bfq2}\hS_{\bfq1\nu}-\delta_{\nu,\bfq1}\hS_{\mu\bfq2}\right.\nn
\left.+\delta_{\mu,\bfq1}\hS_{\bfq2\nu}-\delta_{\nu,\bfq2}\hS_{\mu\bfq1}\right)\nn
+\Oct\sumq\left(\delta_{\mu\bfq4}\hS_{\bfq3\nu}-\delta_{\nu,\bfq3}\hS_{\mu\bfq4}\right.\nn
\left.+\delta_{\mu,\bfq3}\hS_{\bfq4\nu}-\delta_{\nu,\bfq4}\hS_{\mu\bfq3}\right)
\eeqa
At this point, the variables used in the rate equations (\ref{ndy}-\ref{ndy1}), can be precisely defined as
\beqa
N_1&:=&\la \hN_{\bfz1}\ra=\la \hS_{\bfz1\bfz1}\ra,\\
N_2&:=&\la\hN_{\bfz2}\ra=\la\hS_{\bfz2\bfz2}\ra,\\
\varrho_{12}&:=&\la \hS_{\bfz1\bfz2}\ra,\\
N_{\bfq3}&:=&\la\hN_{\bfq3}\ra=\la\hS_{\bfq3\bfq3}\ra,\\
N_{\bfq4}&:=&\la\hN_{\bfq4}\ra=\la\hS_{\bfq4\bfq4}\ra,\\
\varrho_{\bfq34}&=&\la\hS_{\bfq3\bfq4}\ra.
\eeqa
The occupation numbers $N_1$ and $N_2$ count the number of atoms in internal states $|1\ra$ and $|2\ra$, respectively, that have not been `lost'  via emission of a rogue idler photon. The occupation numbers $N_{\bfq3}$ and $N_{\bfq4}$ count the number of atoms in internal states $|3\ra$ and $|4\ra$, which are displaced in momentum space by $\bfK+\bfq$ and $\bfq$, respectively, relative to their initial momenta, corresponding to their having emitted a signal photon with momentum $\bfk\approx -\bfq$. The coherence terms $\varrho_{12}$ and $\varrho_{\bfq34}$ describe the coherence generated by the driving and coupling lasers, respectively.

Beginning with $N_1$, we can derive its equation of motion from (\ref{dSmndt}) by setting $\mu=\nu=\bfz1$. For the commutators (\ref{ckdSmn})-(\ref{HsSmn}) we find
\beqa
\left[\hc_\bfk^\dag,\hS_{\bfz1\bfz1}\right]&=&\sumq f\str(\bfk-\bfq)\hS_{\bfq4\bfz1},\\
\left[\hS_{\bfz1\bfz1},\hc_\bfk\right]&=&\sumq f(\bfk-\bfq)\hS_{\bfz1\bfq4},\\
\left[\hH_s,\hS_{\bfz1\bfz1}\right]&=&\Odt\left(\hS_{\bfz1\bfz2}-\hS_{\bfz2\bfz1}\right),
\eeqa
which leads to
\begin{widetext}
\beq
\ddt N_1=\sumk\pi\delta(\omega_\bfk-\omega_s)|g_\bfk|^2\sum_{\bfq,\bfQ,\bfQ'}f\str(\bfk-\bfq)f(\bfk+\bfQ'-\bfQ)
\left(\la\hS_{\bfq4\bfz1}\hS_{\bfQ'1\bfQ4}\ra+\la\hS_{\bfq4\bfz1}\hS_{\bfQ'3\bfQ2}\ra\right)-i\Odt\varrho_{12}+c.c.
\eeq
\end{widetext}
With the help of (\ref{Smban}) we see that
\beq
\label{dN1dt1}
\la\hS_{\bfq4\bfz1}\hS_{\bfQ'1\bfQ4}\ra=\delta_{\bfQ',0}\la\hS_{\bfq4\bfQ4}\ra+\sum_{{\ss j,J=1}\atop{\ss J\neq j}}^N\la \hS_{\bfq4\bfQ4}\o{j}\hS_{\bfQ'1\bfz1}\o{J}\ra.
\eeq
To implement the approximation that all atoms that emit rogue idler photons are permanently `lost', we simply evaluate expectation values under the assumption that there are no atoms in the states $|u_{\bfq1}\ra$ and $|u_{\bfq2}\ra$ for $\bfq\neq 0$. This allows us to make the simplifications
\beq
\la\hS_{\bfq4\bfz1}\hS_{\bfQ'1\bfQ4}\ra=\delta_{\bfQ',0}\left(N_{\bfq4}+\la \hN_{\bfq4}\hN_{\bfz1}\ra\right)
\eeq
and
\beq
\la\hS_{\bfq4\bfz1}\hS_{\bfQ'3\bfQ2}\ra=\delta_{\bfQ,0}\la\hS_{\bfq4\bfz1}\hS_{\bfQ'3\bfz2}\ra.
\eeq
Inserting these into (\ref{dN1dt1}) and making the approximation
\beq
\label{fapprox}
f\str(\bfk-\bfq)f(\bfk-\bfQ)\approx|f(\bfk-\bfq)|^2\,\delta_{\bfQ,\bfq}\,,
\eeq
then gives
\beqa
\label{dN1dta}
\ddt N_1&=&\frac{\Gamma}{2}\sumq\beta_\bfq\left(\la\hN_{\bfq4}\hN_{\bfz1}\ra+N_{\bfq4}+\la\hS_{\bfq4\bfz1}\hS_{-\bfq3\bfz2}\ra\right)\nn
&-&i\Odt\varrho_{12}+c.c.\, ,
\eeqa
where we have introduced the branching ratio
\beq
\beta_\bfq:=\frac{2}{\Gamma}\sumk\pi\delta(\omega_\bfk-\omega_s)|g_\bfk|^2|f(\bfk-\bfq)|^2,
\eeq
where
\beq
\Gamma=\sum_\bfk 2\pi\delta(\omega_\bfk-\omega_s)|g_\bfk|^2
\eeq
is the spontaneous emission rate for states $|2\ra$ and $|4\ra$, which we have implicitly set equal by making $\hc_\bfk$ (\ref{ck}) symmetric with respect to the
two transitions.

The next step is to show that (a)  $\la \hN_{\bfq4}\hN_{\bfz1}\ra\approx \la\hN_{\bfq4}\ra\la\hN_{\bfz1}\ra$, and (b) that $\la\hS_{\bfq4\bfz1} \hS_{-\bfq3\bfz2}\ra\ll \la\hN_{\bfq4}\hN_{\bfz1}\ra$, so that it can be safely neglected. In order to verify (a) and (b), it is useful to define the excitation-number operators
\beq
\hN_\bfq:=\hN_{\bfq3}+\hN_{\bfq4}.
\eeq
and
\beq
\hN_e:=\sumq\hN_\bfq.
\eeq
with the number of non-excited atoms then given by
\beq
\hN_0:=\hN_1+\hN_2=N-\hN_e
\eeq
In terms of probabilities, the unproven approximation (a) can be re-expressed as
\beq
\label{factorization}
N(N-1)P_2(\bfz1,\bfq4)\approx N^2 P_1(\bfz1)P_1(\bfq4),
\eeq
where $P_2(\mu,\nu)$ is the joint probability that for any ordered pair of atoms, the first one will be in state $|u_\mu\ra$ and the second in state $|u_\nu\ra$, whereas $P_1(\mu)=N_\mu/N$ is the bare probability that any given atom will be in state $|u_\mu\ra$. According to Bayes theorem, we have
\beq
P_2(\bfq4,\bfz1)=P_1(\bfz1|\bfq4)P_1(\bfq4)
\eeq
where $P_1(\bfz1|\bfq4)$ is the conditional probability to find a particular atom in state $|u_{\bfz1}\ra$ given that another particular atom is in state $|u_{\bfq4}\ra$. Knowing that the second atom is in state $|u_{\bfq4}\ra$ means that of the remaining $N-1$ atoms, the average number of excited atoms is now $N_e-1$, or equivalently, out of the remaining $N-1$ atoms, the average number in state $|u_{\bfz1}\ra$ is still $N_1$, Thus we see that
\beq
P_1(\bfz1|\bfq4)=\frac{N_1}{N-1}= P_1(\bfz1)\frac{N}{N-1}.
\eeq
This leads to the result
\beq
\la \hN_{\bfq4}\hN_{\bfz1}\ra=\la \hN_{\bfq4}\ra\la\hN_\bfq1\ra,
\eeq
i.e. the factorization (a) is exact. We note that this would not be obtained without assuming a symmetrized wave-function with fixed total atom number, which allowed us to replace (a) with Eq. (\ref{factorization}).

Turning now to the approximation (b), we begin by introducing the reduced two-body density operator
\beq
\rho_2(\mu,\alpha;\nu,\beta):=\la\hS_{\mu\nu}\hS_{\alpha\beta}\ra,
\eeq
so that $\la \hS_{\bfq4\bfz1}\hS_{-\bfq3\bfz2}\ra=\rho_2(\bfq4,{-}\bfq3;\bfz1,\bfz2)$. If we assume that the probability to have $N_\bfq>1$ is negligible, it follows that the matrix element $\rho_2(\bfq4,{-}\bfq3;\bfz1,\bfz2)$ is a measure of the coherence between the $N_\bfq=N_{-\bfq}=0$ manifold, and the  $N_\bfq=N_{-\bfq}=1$ manifold.

To understand the origin of such coherence in the system, let us start from the initial state
$|\psi_i\ra=\prodj|u_{\bfz1}\ra$, and assume that at time $t=0$, the driving and coupling laser beams are turned on. The atoms will adiabatically follow the ground state of the dressed system, so that immediately after the fields are turned on, the state of the system will be
\beq
|\psi_0\ra=\prodj\left(|u_{\bfz1}\ra\o{j}-i\frac{\Omega_d}{\Gamma}|u_{\bfz2}\ra\o{j}\right).
\eeq
This dressed state will decay by emitting a signal photons at the rate $\Gamma_{eff}=N\Omega_d^2/\Gamma$. The signal photons are distributed over the many recoil modes of the system, with branching ratio $\beta$ per mode. Thus from the perspective of a single $N_\bfq$ manifold, the time delay between signal photons is $T_0=\beta/\Gamma_{err}=\Gamma/(\Omega_d^2 D)$, where we have used $D\approx N\beta$.

 Let us assume that the first photon is emitted along $\bfk\approx-\bfq$,  causing the system to jump from the $N_\bfq=N_{-\bfq}=0$ manifold to the $N_\bfq=1$, $N_{-\bfq}=0$ manifold. The normalized state immediately after this quantum jump is given by
\beq
|\psi_1\ra=\frac{\hc_{-\bfq}|\psi_0\ra}{\la\psi_0|\hc^\dag_{-\bfq}\hc_{-\bfq}|\psi_0\ra^{1/2}}.
\eeq
By neglecting rogue-photon emission and making the approximation (\ref{fapprox}), we can simplify (\ref{cku}) to
\beq
\label{cq}
\hc_\bfq=\hS_{\bfz1\bfq4}+\hS_{-\bfq3\bfz2},
\eeq
which leads to
\beq
|\psi_1\ra=-i\frac{\Gamma}{\Omega_d N}\hS_{\bfq3\bfz2}|\psi_0\ra.
\eeq
This state will live for time $t_1\sim T_e=\frac{1}{\Gamma D}$, after which a second photon will be emitted along the $\bfk\approx \bfq$ direction. With $t=0$ corresponding to the emission of the second photon, the state of the system at time $t$ later is then
\beq
|\psi_2(t)\ra=\frac{U_c(t)\hc_{\bfq}U_c(t_1)|\psi_1\ra}{\la\psi_1|U_c^\dag(t_1)\hc^\dag_\bfq\hc_\bfq U_c(t_1)|\psi_1\ra^{1/2}},
\eeq
where the propagator
\beq
U_c(t)=\exp\left[-i\frac{\Omega_ct}{2}\sum_\bfq\left(\hS_{\bfq3\bfq4}+\hS_{\bfq4\bfq3}\right)\right],
\eeq
describes the Rabi oscillations between states $|3\ra$ and $|4\ra$, that occur when $\Omega_c>\Gamma D/2$.

We can take $U_c(t_1)\hS_{\bfq3\bfz2}\approx \hS_{\bfq4\bfz2}$, as $U_c(t)$ mixes levels $|3\ra$ and $|4\ra$, and the highest probability of emission occurs when the excited atom is in level $|4\ra$. This leads to the result
\beq
|\psi_2(t)\ra\approx\frac{\Gamma}{\Omega_dN}\left(\hS_{\bfz1\bfz2}+U_c(t)\hS_{-\bfq3\bfz2}\hS_{\bfq4\bfz2}\right)|\psi_0\ra,
\eeq
which shows that the state $|\psi_2(t)\ra$ is in fact a coherent superposition of a state with $N_\bfq=N_{-\bfq}=0$, corresponding to the second photon being an idler photon (emitted on the $|4\ra\to|1\ra$ transition) and a state with $N_\bfq=N_{-\bfq}=1$, corresponding to the second photon being a new signal photon (emitted on the $|2\ra\to|3\ra$ transition). The lifetime of this coherent superposition state is $T_{coh}\sim T_e$, the timescale on which either (i) a second pair of photons will be emitted, confirming that the second photon was a signal photon, or (ii) no additional photons will be emitted, confirming that the second photon was the idler photon. In either case, the system will collapse back onto a state with $N_\bfq=N_{-\bfq}=0$.

From this analysis, we now see that
\beqa
\la \hS_{\bfq4\bfz1}\hS_{-\bfq3\bfz2}\ra&\approx&\frac{T_e}{T_0}\la\psi_2(t)|\hS_{\bfq4\bfz1}\hS_{-\bfq3\bfz2}|\psi_2(t)\ra\nn
&\approx&\cos^2(\Omega_ct/2)\frac{\Omega_d^4}{\Gamma^4}N
\eeqa
where the factor $T_e/T_0=\Omega_d^2/\Gamma^2$ is the probability to find the system in state $|\psi_2(t)\ra$. To compare this to the term $\la \hN_{\bfq4}\hN_{\bfz1}\ra=N^2P_1(\bfz1)P_1(\bfq4)$ in (\ref{dN1dta}), we need to estimate the single particle probabilities, $P_1(\mu)$.
Estimates for these probabilites can be found from the equilibrium condition $R_s=R_I$.  For $N_e\ll N$ and $\Omega_c\ge \Gamma D$,  we can make the simple estimates $R_s=N_2\Gamma$ and $R_I=\frac{1}{2}N_e \Gamma D$. With $N_2=\frac{\Omega_d^2}{\Gamma^2}N_1$ and $N_e=1-N_0$, we find $N_1=\frac{\Gamma^2N}{\Gamma^2+\Omega_d^2(1+2/D)}$, $N_2=\frac{\Omega_d^2N}{\Gamma^2+\Omega_d^2(1+2/D)}$, $N_e=\frac{(2/D)\Omega_d^2N}{\Gamma^2+\Omega_d^2(1+2/D)}$, and  $N_{\bfq3}\approx N_{\bfq4}\approx \frac{1}{2}\beta N_e$, which gives us
\beq
P_1(\bfz1)\approx 1-\frac{\Omega_d^2}{\Gamma^2},
\eeq
and
\beq
P_1(\bfq4)\approx\frac{\Omega_d^2}{\Gamma^2}\frac{1}{N}.
\eeq
Thus we see that
\beq
\la \hN_{\bfq4}\hN_{\bfz1}\ra\approx \frac{\Omega_d^2}{\Gamma^2}N,
\eeq
which is larger than $\la \hS_{\bfq4\bfz1}\hS_{-\bfq3\bfz1}\ra$ by a factor $(\Gamma/\Omega_d)^2$. For the parameters used in the numerical simulations of Sec. (\ref{Sec.BM.PD}), this is a factor of $100$. Keeping only the dominant term, $\la \hN_{\bfq4}\hN_{\bfz1}\ra$, then leads to
\beq
\ddt N_1=\Gamma\sumq\beta_\bfq N_{\bfq4}(N_1+1)-i\frac{\Omega_d}{2}\left(\varrho_{12}-c.c.\right),
\eeq
which for $\Gamma_4=\Gamma$ and $\beta_{\bfq4}=\beta_\bfq$, validates the rate equation (\ref{ndy}). Using the general methods outlined in this appendix, the remaining rate equations (\ref{ndy1a})-(\ref{ndy1}) can be derived as well. Clearly deriving these rate equations from first principles is highly non-trivial. However, once their validity is established, the fact that they follow an established form allows one to write them down directly, rather than re-derive them for each particular model.

\section{Extended Rate equations}
\label{Sec.APP.ERE}

In this section, we give the extended rate equations we have used to
model the Cesium scheme presented in section \ref{Sec.RM.CA}. In the
scheme, driving from $|1\rangle$ to $|2\rangle$ is accomplished via
a three-photon process, through detuned intermediate levels of
$|a\rangle\equiv |6P_{3/2},m_F=3\rangle$ and $|b\rangle\equiv
|6D_{5/2}, m_F=4\rangle$, as shown in figure \ref{fig5} (a). With
the conventions of $N_j$ being atom number in state $|j\rangle$ and
$\varrho_{jh}$ being the coherence between $|j\rangle$ and
$|h\rangle$, the equations are obtained as
\begin{widetext}
 \begin{eqnarray}
    \label{ndy2}
    \ddt N_1 &=& \frac{i}{2}\left(\Omega_1\varrho_{a1}-c.c\right)-\Gamma_1 N_1+\alpha_{41}\Gamma^0_4\sum_\bfk
             \beta_{\bfk 4} N_{\bfk 4} (N_1+1), \\
    \ddt N_a &=& -\frac{i}{2}\left(\Omega_1\varrho_{a1}-c.c\right)
            +\frac{i}{2}\left(\Omega_2\varrho_{ba}-c.c\right)-\Gamma^0_a N_a, \\
    \ddt N_b &=& -\frac{i}{2}\left(\Omega_2\varrho_{ba}-c.c\right)
         -\frac{i}{2}\left(\Omega_3\varrho_{b2}-c.c\right)-\Gamma^0_b N_b, \\
    \ddt N_2 &=& \frac{i}{2}\left(\Omega_3\varrho_{b2}-c.c\right)-\Gamma_2\sum_\bfk
             \beta_{\bfk 2} N_2 (N_{\bfk 3}+1), \\
    \ddt \varrho_{a1} &=& i\frac{\Omega_1}{2}(N_1-N_a)+i\frac{\Omega_2}{2}\rho_{b1}+\left(i\Delta_1
             -\frac{1}{2}(\Gamma^0_a+\Gamma_1)+\frac{1}{2}\alpha_{41}\Gamma^0_4\sum_\bfk \beta_{\bfk 4} N_{\bfk 4}\right)\varrho_{a1}
              \\
    \ddt \varrho_{ba} &=& i\frac{\Omega_2}{2}(N_a-N_b)-i\frac{\Omega_1}{2}\varrho_{b1}+i\frac{\Omega_3}{2}\varrho^\ast_{a2}
         -\left(i(\Delta_2+\Delta_1)+\frac{1}{2}(\Gamma^0_a+\Gamma^0_b)\right)\varrho_{ba} \\
    \ddt \varrho_{b1}&=& i\frac{\Omega_2}{2}\rho_{a1}-i\frac{\Omega_1}{2}\varrho_{ba}+i\frac{\Omega_3}{2}
        \varrho_{21}+\left(\frac{1}{2}\alpha_{41}\Gamma^0_4\sum_\bfk \beta_{\bfk 4} N_{\bfk 4}-\frac{1}{2}(\Gamma_1+\Gamma^0_b)\right)\varrho_{b1}
        +i\Delta_2\varrho_{b1}, \\
   \ddt \varrho_{b2}&=& i\frac{\Omega_3}{2}
        (N_2-N_b)+i\frac{\Omega_2}{2}\varrho_{a2}+\left(i(\Delta_3-\Delta_2)-\frac{\Gamma^0_b}{2}
        -\frac{1}{2}\Gamma_2 \sum_\bfk\beta_{\bfk 2} (N_{\bfk 3}+1)\right)\varrho_{b2}, \\
    \ddt \varrho_{a2}&=& i\frac{\Omega_1}{2}\rho_{12}+i\frac{\Omega_2}{2}\varrho_{b2}
           -i\frac{\Omega_3}{2}\varrho^\ast_{ba}+\left(i(\Delta_1+\Delta_3)-\frac{1}{2}\Gamma^0_a
         -\sum_\bfk\Gamma_2 \beta_{\bfk 2} (N_{\bfk 3}+1)\right) \varrho_{a2}\\
    \ddt \varrho_{21}&=& -i\frac{\Omega_1}{2}\rho^\ast_{a2}+i\frac{\Omega^\ast_3}{2}\varrho_{b1}
        +\left[i\Delta_3-\frac{1}{2}\Gamma_1-\sum_\bfk \Big(\Gamma_2 \beta_{\bfk 2} (N_{\bfk 3}+1)
        -\alpha_{41}\Gamma^0_4 \beta_{\bfk 4} N_{\bfk 4} \Big) \right]\varrho_{21}\\
    \ddt N_{\bfk 3} &=& \frac{i}{2} \left(\Omega_4\varrho_{\bfk 43}-c.c\right)+
            \Gamma_2 \beta_{\bfk 2} N_2 (N_{\bfk 3}+1)-\Gamma_3 N_3, \\
    \ddt N_{\bfk 4} &=& -\frac{i}{2}\left(\Omega_4 \varrho_{\bfk 43}-c.c\right)-
             \Gamma^0_4 N_{\bfk 4} ( \alpha_{41}\beta_{\bfk 4}N_1+1), \\
    \ddt \varrho_{\bfk 43} &=& i \frac{\Omega_4}{2} (N_{\bfk 3}-N_{\bfk 4})+\frac{1}{2}
      \bigg(\Gamma_2\mu_{\bfk 2} N_2-\Gamma_3-\Gamma^0_4(\alpha_{41} \mu_{\bfk 4} N_1+1) \bigg)\varrho_{\bfk 43},
\end{eqnarray}
\end{widetext}
assuming all $\Omega's$ are real. Here, $\Delta_j$ ($j=1,2,3$) is
the detuning indicated in figure \ref{fig5} (a). $\Gamma^0_{a}$,
$\Gamma^0_{b}$, $\Gamma_2$ and $\Gamma^0_4$ are the natural
linewidths of states $|a\rangle$, $|b\rangle$, $|2\rangle$ and
$|4\rangle$, respectively. $\alpha_{41}$ is the branch percentage
for $|4\rangle$ to spontaneously decay to $|1\rangle$, so that
$\Gamma_4=\alpha_{41}\Gamma^0_4$ is the spontaneous decay rate from
$|4\rangle$ to $|1\rangle$. $\Gamma_1=\frac{\Omega_4^2
}{4\Delta^2_3}\Gamma_a$ is the effective loss rate of $|1\rangle$
atoms, due to being excited by the detuned laser $\Omega_4$ to the
unstable upper level $|a\rangle$. Similarly, $\Gamma_3=
\frac{\Omega^2_1}{4(\Delta_1-\Delta_3)^2}\Gamma^0_4$ is the
effective loss rate for atoms in $|3\rangle$, induced by laser
$\Omega_1$. We note the above rate-equation model has only
incorporated dominant atom loss mechanism for each atomic level.

\end{appendix}

\end{document}